\documentclass[12pt,preprint]{aastex}
\usepackage{amssymb}

\newcommand{\HII}{\mbox{H\thinspace{\sc ii}}}  %
\newcommand{\wtw}{W3(H$_2$O)} %
\newcommand{\Msun}{\mbox{~M$_{\odot}$}} %
\newcommand{\Lsun}{\mbox{~L$_{\odot}$}} %

\begin{document}
\title{A High-Mass Protobinary System in the Hot Core W3(H$_2$O)}
\author{Huei-Ru Chen\altaffilmark{1}, William J. Welch\altaffilmark{2}, 
David J. Wilner\altaffilmark{3}, \& Edmund C. Sutton\altaffilmark{4}}
\altaffiltext{1}{Institute of Astronomy and Astrophysics, Academia Sinica, P.O. Box 23141, Taipei 106, Taiwan; hchen@asiaa.sinica.edu.tw }
\altaffiltext{2}{Department of Astronomy, University of California, Berkeley, CA 94720; welch@astro.berkeley.edu}
\altaffiltext{3}{Harvard-Smithsonian Center for Astrophysics, 60 Garden Street, Cambridge, MA 02138; dwilner@cfa.harvard.edu}
\altaffiltext{4}{Department of Astronomy, University of Illinois, 1002 West Green Street, Urbana, IL 61801; sutton@astro.uiuc.edu}

\begin{abstract} 
We have observed a high-mass protobinary system in the hot core \wtw\ with the BIMA Array.  
Our continuum maps at wavelengths of 1.4~mm and 2.8~mm both achieve 
sub-arcsecond angular resolutions and show a double-peaked morphology.
The angular separation of the two sources is 1\farcs19 corresponding to
$2.43 \times 10^3$~AU at the source distance of 2.04~kpc.
The flux densities of the two sources at 1.4~mm and 2.8~mm have a spectral index of 3, 
translating to an opacity law of $\kappa_{\nu} \propto \nu$.
The small spectral indices suggest that grain growth has begun in the hot core.
We have also observed 5 $K$ components of the methyl cyanide 
($\mathrm{CH_3CN}$) $J=12 \rightarrow 11$ transitions.  
A radial velocity difference of $2.81 \pm 0.10 \; \mathrm{km \, s^{-1}}$ is found towards the two
continuum peaks.  
Interpreting these two sources as binary components in orbit about one another, 
we find a minimum mass of 22\Msun\ for the system.
Radiative transfer models are constructed to explain both the continuum and 
methyl cyanide line observations of each source. 
Power-law distributions of both density and temperature are derived.  
Density distributions close to the free-fall value, $r^{-1.5}$, are found for both 
components, suggesting continuing accretion.
The derived luminosities suggest the two sources have equivalent zero-age main sequence 
(ZAMS) spectral type B0.5 - B0.
The nebular masses derived from the continuum observations are about 5\Msun\ for 
source~A and 4\Msun\ for source~C.
A velocity gradient previously detected may be explained by 
unresolved binary rotation with a small velocity difference.
\end{abstract}

\section{INTRODUCTION}
Massive star formation presents a challenge in both observational and theoretical studies.
Unlike their low-mass counterparts, massive young stellar objects (YSOs) are typically 
distant ($d \gtrsim 0.5$~kpc), obscured ($A_V \gtrsim 100$), and often observed in 
clustered environments.
With the development of millimeter interferometers, it has become possible to resolve 
individual, deeply embedded, massive YSOs.  
In the past two decades, a few warm ($T \gtrsim 100$~K), dense ($n \gtrsim 10^6$~cm$^{-3}$), 
molecular clumps of high luminosities ($L \simeq 10^4 - 10^5 \Lsun$) have been discovered 
in the vicinity of ultracompact (UC) \HII\ regions \markcite{Turner84,Kurtz00}({Turner} \& {Welch} 1984; {Kurtz} {et~al.} 2000).
These molecular clumps exhibit chemistry that is characterized by 
high abundances of fully hydrogenated species \markcite{vanDishoeck98}({van Dishoeck} \& {Blake} 1998), including 
a variety of organic molecules, such as methyl cyanide ($\mathrm{CH_3CN}$), 
ethanol ($\mathrm{C_2H_5OH}$), etc.
In a few cases, internal heating of the clumps is showed directly by radial temperature 
gradients with ammonia ($\mathrm{NH_3}$) emission \markcite{Cesaroni98}({Cesaroni} {et~al.} 1998).
These molecular clumps were later classified as hot cores.
The lack of detectable Str\"{o}mgren spheres further places hot cores in 
an evolutionary phase earlier than UC \HII\ regions \markcite{Wilner01}({Wilner} {et~al.} 2001). 
Maser emission, such as water and methanol masers, is often associated with hot cores and 
suggestive of violent collision events \markcite{Tarter86}({Tarter} \& {Welch} 1986).
One most interesting phenomenon is an apparent velocity gradient that has been detected in a
few hot cores, for example, the Orion hot core, \wtw, IRAS~20126+4104, NGC~7538~S,  
G29.96-0.02, G31.41+0.31, and IRAS~07427-2400 
\markcite{Plambeck90,Wyrowski97,Cesaroni99,Sandell03,Olmi03,Cesaroni94b,Kumar03}({Plambeck}, {Wright}, \&  {Carlstrom} 1990; {Wyrowski} {et~al.} 1997; {Cesaroni} {et~al.} 1999; {Sandell}, {Wright}, \&  {Forster} 2003; {Olmi} {et~al.} 2003; {Cesaroni} {et~al.} 1994; {Kumar} {et~al.} 2003). 
Although the emission is spatially unresolved in each velocity channel, 
the velocity gradient has been long proposed as evidence for the Keplerian rotation of 
a circumstellar disk.

At a distance of 2.04~kpc \markcite{Hachisuka06}({Hachisuka} {et~al.} 2006), 
the neighborhood of the W3(OH) ultracompact (UC) \HII\ region is a nearby, well-studied star-forming region in the 
course of developing an OB stellar group.
The most recent star-forming activity is taking place in a hot molecular core, \wtw, 
which appears as a warm, dense molecular clump 6\arcsec\ to the east of W3(OH).
This hot core was first discovered by \markcite{Turner84}{Turner} \& {Welch} (1984) 
through an aperture synthesis study of HCN ($1 \rightarrow 0$) emission.
The broad line width observed in HCN indicated the presence of an obscured young stellar object, 
possibly a protostar, with an estimated luminosity of $10^4$\Lsun.
Proper motions of water masers in \wtw\ can be explained as part of a bipolar outflow 
\markcite{Alcolea93}({Alcolea} {et~al.} 1993).
\markcite{Wilner95}{Wilner}, {Welch}, \& {Forster} (1995) detected unresolved dust emission at 87.7~GHz with a flux density of 43~mJy, 
implying a mass of 10-20\Msun\ and a luminosity of $10^3$-$10^4$\Lsun.
At centimeter wavelengths, observations with the Very Large Array (VLA) revealed a synchrotron jet 
near the expansion center of the water masers \markcite{Reid95,Wilner99}({Reid} {et~al.} 1995; {Wilner}, {Reid}, \& {Menten} 1999).
A secondary continuum source with a rising spectral index was also detected, possibly featuring an additional young stellar object \markcite{Wilner99}({Wilner} {et~al.} 1999). 
Subsequent line and continuum observations at 220~GHz \markcite{Wyrowski99}({Wyrowski} {et~al.} 1999) resolved 
the hot core into three emission components, with two components (A and C) roughly coincident
with two temperature peaks of about 200~K, derived from the HNCO $K_a = 0, 2,$ and $3$ lines.
In addition, a linear velocity gradient spanning about \mbox{$10 \; \mathrm{km \, s^{-1}}$} 
along the East-West direction was found within a region of 1\arcsec\ across 
the hot core \markcite{Wyrowski97}({Wyrowski} {et~al.} 1997).
This gradient was found by plotting the positions of maximum
emission derived from channel maps in several lines.  The total gradient
spanned an angular range of about 1\arcsec, smaller than the instrumental resolution.
Thus, it was unclear what could be the source of the gradient.  
\markcite{Wyrowski97}{Wyrowski} {et~al.} (1997) proposed three interesting possibilities.  
The first suggestion was that the high excitation molecules traced the sides of 
the cavity carved out by the outflows associated with the water masers.  
The second idea was that the gradient represents a large disk seen edge-on.  
The third idea was that there might be a double source with different radial 
velocities with the components associated with the two centimeter 
continuum sources.

The present study is an extension of the Plateau de Bure work 
\markcite{Wyrowski97,Wyrowski99}({Wyrowski} {et~al.} 1997, 1999) enabled by the Berkeley-Illinois-Maryland Association (BIMA) Millimeter Array with its longer 
baselines and more extensive $uv$ coverage.  
Although the BIMA Array has less collecting area, the \wtw\ source is sufficiently 
bright to be imaged.  
We have mapped the continuum emission at wavelengths of 1.4~mm and 
2.8~mm at very high angular resolution in search of the large disk or whatever else 
might be the source of the velocity gradient.  
The angular resolution is 0\farcs26 at 1.4~mm and 0\farcs40 at 2.8~mm, 
allowing us to study the structure and dust properties of the components within 
the hot core in detail.
We have also observed the $K=2,3,4,5,$ and $6$ components of the 
$\mathrm{CH_3CN}$ $J=12 \rightarrow 11$ transitions
with 1\arcsec\ resolution to derive the temperature at each continuum peak.
The kinematics are also obtained with high precision by simultaneously fitting 
Gaussians to each of the 5 observed $K$ components. 

\section{OBSERVATIONS AND DATA REDUCTION}
The continuum emission was mapped at both 1.4~mm and 2.8~mm in the A and B 
configurations of the BIMA Array.
The pointing center was ($\alpha$,$\delta$)(J2000) = 
($2^{\rm h}27^{\rm m}3\fs87,+61\arcdeg52\arcmin24\farcs6$). 
The calibrators were 0359+509 at 1.4~mm and 0228+673 at 2.8~mm.
The flux densities of the calibrators were determined by comparison with 
the Be star MWC~349, whose flux densities were assumed to be 1.2~Jy and 1.9~Jy 
at 2.8~mm and 1.4~mm, respectively 
\markcite{Dreher83,Tafoya04}({Dreher} \& {Welch} 1983; {{Tafoya}, {G{\' o}mez}, \& {Rodr{\'{\i}}guez}} 2004).
The accuracy of the flux scale was estimated with the variance in the flux measurements 
of 3C84, and should be good within 15\%.

The 1.4~mm data set taken in the A configuration provided highest spatial frequencies, which
resolved the UC \HII\ region W3(OH) into a shell-like structure.
The three continuum data sets at 1.4~mm were observed in fast-switching mode, 
consisting of a 2 minute cycle on 0359+509, 0228+673, and W3(OH) to trace
the rapid atmospheric phase fluctuation.
Additionally, a 1.5 minute integration on 0359+509 was repeated every 20 minutes
to help gain amplitude solutions.
The reference source, 0228+673, was included to monitor the imaging quality.
The typical system temperature was $400 \sim 600$~K, and the typical phase noise 
was about $\phi_{\mathrm{rms}} \simeq 35\arcdeg$.
The 2.8~mm continuum data sets were observed in a regular fashion, which repeats
the calibrator, 0228+673, every 5 minutes. 

The calibration and imaging were performed with the MIRIAD software package.
The two sidebands were calibrated independently.
Every data set was first calibrated by deriving phase-only gain solutions from the 
calibrator on a timescale of 2 or 5 minutes, followed by solving complex gains on 
a longer timescale of $15$ to $20$ minutes.
The combined complex gains were transfered to the source data, which were further 
improved with one iteration of phase-only self calibration, with the exception of the 
1.4~mm A-configuration data that showed no improvement after self calibration.
The self-calibration intervals were about 1 or 3 minutes, depending on 
the wavelengths and configurations.
The self calibration was performed on W3(OH) for the A-configuration data and on both 
W3(OH) and \wtw\ for the B-configuration data.
When multiple data sets were available, a combined map was made to be the
model for self calibration.
The astrometery at 1.4~mm after the self-calibration was checked with reference
quasar, 0228+673: first applied the phase-only gain solutions to 0228+673, 
generated an image, and then performed Gaussian fit to find the emission centroid.
The positional errors were less than 0\farcs014 and 0\farcs05 for the 
A- and B-configuration data, respectively.
Because the observations in the A~configuration resolved \wtw, the images were formed by 
combining the visibilities from both the A and B configurations.  
Both sidebands were used in a multi-frequency synthesis to generate line-free continuum maps.
Each map was then CLEANed in the usual way and was restored with a circular 
Gaussian beam of the same solid angle as its synthesized beam.
Details of  the continuum observations are listed in Table~\ref{contobs}.
\begin{deluxetable}{lcc}
\tablewidth{0pt}
\tablecolumns{3}
\tablecaption{Parameters for the Continuum Observations \label{contobs}}
\tablehead{ \colhead{Parameter} & \colhead{1.4~mm} & \colhead{2.8~mm} }
\startdata
Synthesized frequency & 221.2~GHz & 108.6~GHz \\
Dates (configurations) & 2000 Dec 26 (A) & 2003 Jan 8 (A) \\
                                    &  2001 Jan 31 (B) & 2003 Jan 16 (A) \\
                                    &  2001 Feb 8 (B)   & 2003 Feb 2 (B) \\  
Calibrator & 0359+509 & 0228+673 \\
$uv$-range (k$\lambda$) & 53 - 1000 (A) & 16 - 460 (A) \\
                                         & 10 - 170 (B) & 4 - 85 (B)  \\
Largest visible scale & 2\arcsec (A) \& 10\arcsec (B) & 6\arcsec (A) \& 25\arcsec (B) \\
W3(OH) flux density & 3.41~Jy & 3.61~Jy \\
Primary beam & 52\arcsec & 106\arcsec \\
Weight Scheme & uniform & uniform \\
Synthesized beam & $0\farcs31 \times 0\farcs22$ (77\arcdeg) 
                              & $0\farcs46 \times 0\farcs35$ (72\arcdeg) \\
Restored beam & 0\farcs26 & 0\farcs40 \\
Bandwidth (each sideband) & 800~MHz & 400~MHz \\
Rms noise (mJy/Beam) & 3.0 & 1.7
\enddata
\end{deluxetable}

We have also observed the $K = 2,3,4,5,$ and 6 components of the $\mathrm{CH_3CN}$ 
$J=12 \rightarrow 11$ transitions in the B configuration during the spring 
season of 2003.  
In order to improve the single-to-noise ratio (SNR) on the calibrator, 0359+509, 
for each 5 minutes interval, the correlators were adjusted to include two wide-band 
windows, which provided 400~MHz continuum bandwidth.
However, this correlator setup allowed simultaneous integration of only two high spectral 
resolution windows.
The $K=2,3,$ and $4$ components were observed on February 8 while the $K = 5$ and $6$ 
components were observed on February 9 under nearly identical weather conditions.
The data were reduced with a procedure similar to that of the continuum data.
The velocity resolution of the spectral line observations is $0.53 \; \mathrm{km \, s^{-1}}$. 
The line observation parameters are listed in Table~\ref{mcobs}.
\begin{deluxetable}{lcc}
\tablewidth{0pt}
\tablecolumns{3}
\tablecaption{Parameters for the $\mathrm{CH_3CN}$ Line Observations \label{mcobs}}
\tablehead{  \colhead{Parameters} & \colhead{2003 Feb 08} & \colhead{2003 Feb 09} }
\startdata
Configuration & B  & B \\
Calibrator & 0359+509 & 0359+509 \\
Observed $K$ components &  2,3,4 & 5,6 \\
$uv$ range (k$\lambda$) & 14 - 175 & 14 - 175 \\
Largest visible scale & 7\arcsec & 7\arcsec \\
W3(OH) flux density (Jy) & 3.57 & 3.45 \\
Synthesized beam & $1\farcs2 \times 0\farcs9$ (-62\arcdeg)
                               & $1\farcs2 \times 0\farcs9$ (-63\arcdeg) \\
Restored beam & 1\farcs0 & 1\farcs0 \\
Channel width & $0.53 \; \mathrm{km \, s^{-1}}$ & $0.53 \; \mathrm{km \, s^{-1}}$ \\
Channel rms                        & 2.9~K ($K=2,3$) & 3.0~K ($K=5$) \\
                                            & 2.6~K ($K=4$)  & 3.0~K ($K=6$)
\enddata
\end{deluxetable}

\section{RESULTS}
\subsection{Continuum emission}
\begin{figure}
\plotone{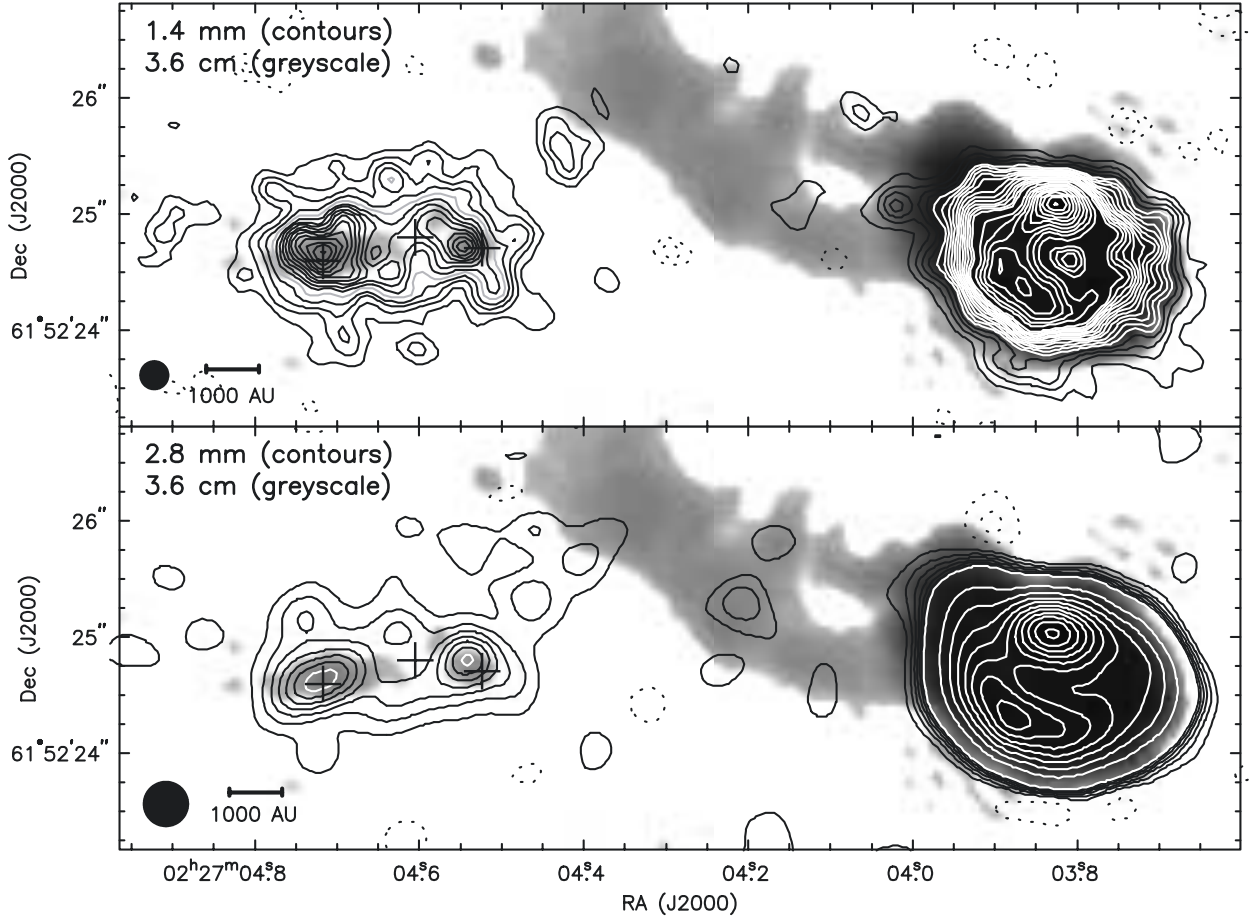}
\caption{BIMA high-resolution line-free continuum maps at 1.4~mm (contours, top panel) and 2.8~mm (contours, bottom panel) overlaid on a 3.6~cm continuum map (greyscale, \markcite{Wilner99}{Wilner} {et~al.} (1999)).
The brightest source is the ultracompact \HII\ region, W3(OH).   
The hot core \wtw, located 6\arcsec\ east to W3(OH), shows a double-peaked morphology 
at both 1.4~mm and 2.8~mm with peak positions coinciding well with those in the 3.6~cm map.
The three crosses indicate the positions of the three continuum peaks, A, B, and C, 
from East to West, observed by \markcite{Wyrowski99}{Wyrowski} {et~al.} (1999). 
The small bar at lower-left corner indicates $10^3$~AU.  
The beam size is 0\farcs26 for the 1.4~mm map and 0\farcs40 for the 2.8~mm map.
Contour levels of the 1.4~mm map correspond to -12, -6, 6, 12 to 96 by 6 ($2 \sigma$), 
and 96 to 222 by 12 ($4 \sigma$)~mJy/Beam.
The white contours in W3(OH) start at 30~mJy ($10 \sigma$), which is described 
by light grey contours in W3(H$_2$O).
Contours of the 2.8~mm map correspond to -6.8, -3.4, 3.4 to 23.8 by 3.4 ($2\sigma$), 
and 23.8 to 533.8 by 51 ($30\sigma$)~mJy/Beam.
The white contours start at 23.8~mJy ($14 \sigma$), which is the highest contour level in 
W3(H$_2$O).
\label{contmap}}
\end{figure}

Fig.~\ref{contmap} shows the millimeter continuum emission (contours) for 
both 1.4~mm and 2.8~mm in the central part of the W3(OH) region overlaid on 
a 3.6~cm continuum map \markcite{Wilner99}({Wilner} {et~al.} 1999, greyscale). 
The brightest source is the UC \HII\ region W3(OH), whose champagne flow to
the northeast is visible at 3.6~cm. 
The hot core \wtw, 6\arcsec\ east to W3(OH), is detected at both 1.4~mm and 2.8~mm
with flux densities of 1.38~Jy and 0.22~Jy, respectively.
The dust emission is much stronger at 1.4~mm.
In addition, a thin layer of dust may still wrap around the UC~\HII\ region and contributes to
the weak emission feature along the edge of W3(OH) in the 1.4~mm map.
We do not detect other sources of continuum emission within the field of view.

For the components in the hot core, we adopt the notation of 
\markcite{Wyrowski99}{Wyrowski} {et~al.} (1999), who identified three peaks: A, B, and C.  
The principal features are A and C in our map.
This is the first time that the separate source C has been imaged at 2.8 mm.    
The B feature is roughly the northeast extension of peak~C in our 1.4~mm map.  
The 1.4~mm map of Fig.~\ref{contmap} agrees best with the 3.6~cm map 
\markcite{Wilner99}({Wilner} {et~al.} 1999).
The peak positions are at offset A = (5\farcs94, 0\farcs05) and 
C = (4\farcs75, 0\farcs10) relative to the pointing center.  
This gives a projected separation of 1\farcs19, equivalent to 
$2.43 \times 10^3 \; \mathrm{AU}$.  

The hot core features of Fig.~\ref{contmap} are best interpreted as two separate
sources, possibly a binary system.  
Source B may be a separate companion to source~C.  
In any case, a large disk seems very unlikely.  
There is very little emission bridging sources A and C, not even enough to 
fit an edge-on doughnut. 
Thus, we adopt the third idea of \markcite{Wyrowski97}{Wyrowski} {et~al.} (1997) that the structure of the hot core 
corresponds to two principal components, and we develop our models for the region 
with the idea that we are observing a protobinary system.  
In Sect.~\ref{sec_cont} and \ref{sec_line}, we show how this two component model nicely explains 
the large velocity gradient discovered by \markcite{Wyrowski97}{Wyrowski} {et~al.} (1997).  
Indeed the small deviations from the straight line that are evident in our data are 
difficult to explain with any other model.

\subsection{Methyl cyanide line emission}
\begin{figure}
\plotone{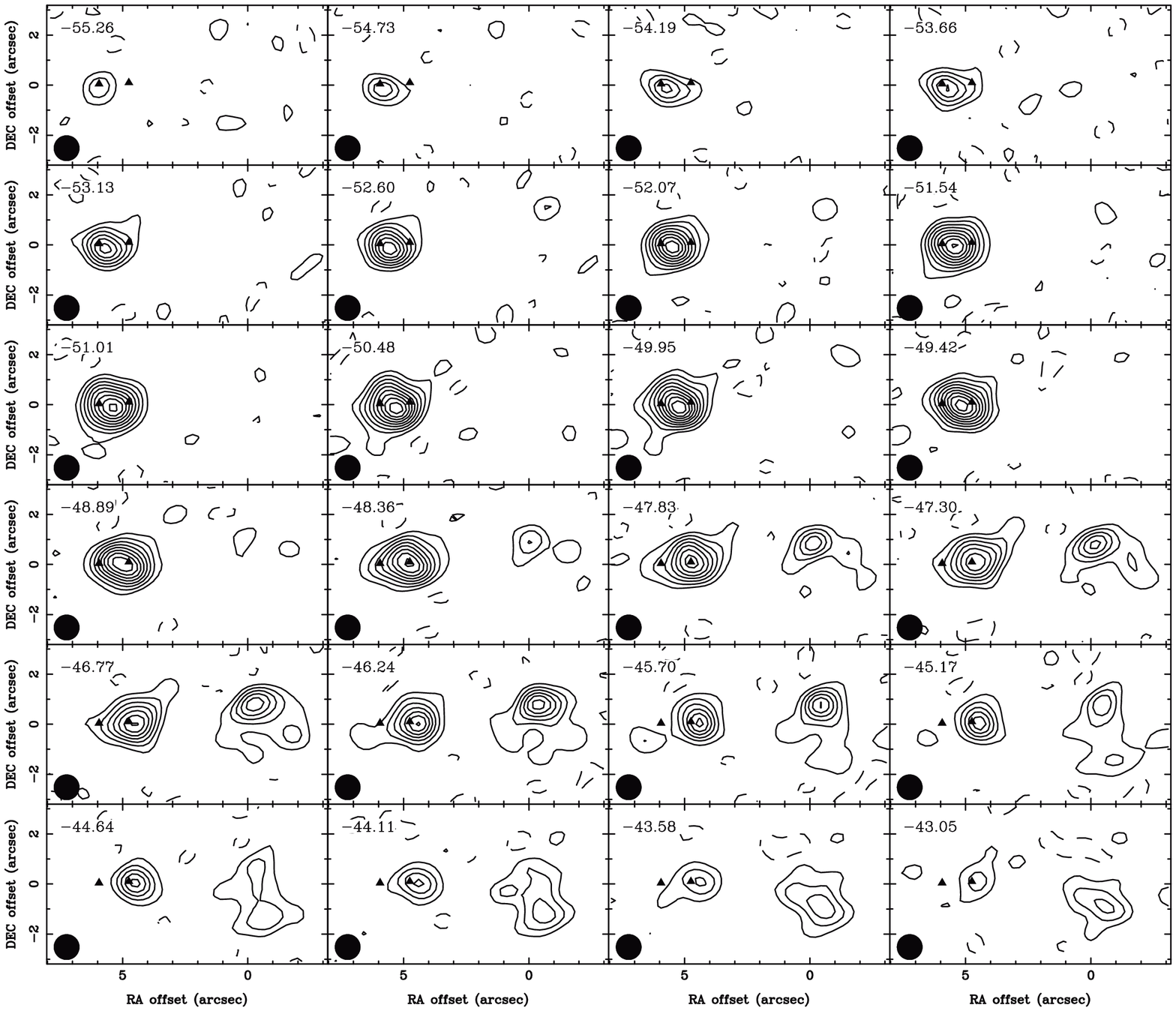}
\caption{
Channel maps of the $K = 3$ component of the $\mathrm{CH_3CN}$ $J=12 \rightarrow 11$
transition.
The two triangles indicate the two continuum peaks, A and C.
In the hot core, the centroid of the emission clearly shows a position drift from East to
West with increasing radial velocity.
The UC \HII\ region, W3(OH), is centered at (0,0) with the $\mathrm{CH_3CN}$ 
emission between $ \upsilon_r = -48$ to $-43 \; \mathrm{km \, s^{-1}}$.
Contour levels start from -5.7, 5.7, 10.4 to 57 by 5.7~K ($2\sigma$).
\label{mcmap}}
\end{figure}

The $K = 2,3,4,5,$ and $6$ components of the $\mathrm{CH_3CN}$ $J=12 \rightarrow 11$ 
transition were observed with a 1\arcsec\ beam. 
The parameters of the observed $K$ components are listed in Table~\ref{obsK}.
We did not observe the $K=0$ and $1$ components because their frequency separation 
of 4.25~MHz is not sufficiently large to avoid overlap with each other.
The large line width expected in star-forming regions ($\gtrsim 2$~MHz in the case of \wtw) 
makes these two components largely unusable for most purposes. 
Fig.~\ref{mcmap} shows the channel maps of the $K = 3$ component.
The UC \HII\ region, W3(OH), centered at (0,0), also shows 
methyl cyanide emission between $\upsilon_r = -49$ to $-45 \; \mathrm{km \, s^{-1}}$. 
The two triangles indicate the positions of the two continuum peaks, A and C.
Although each of the component A and C remains unresolved, 
the hot core is well resolved along the East-West direction.
The centroids of the methyl cyanide emission show a position drift from 
East to West with increasing radial velocity. 
The angular resolution of 1\arcsec\ is smaller than the separation of 1\farcs19 
between source A and C.
This allows us to obtain the spectrum for one source without much interference 
from the other source.
The properties of methyl cyanide molecules are discussed in Appendix~\ref{sec_pmc}.

\begin{deluxetable}{ccccrrc}
\tablewidth{0pt}
\tablecolumns{7}
\tablecaption{Observed $K$ Components of $\mathrm{CH_3CN}$ ($J = 12-11$) Transitions \label{obsK} }
\tablehead{
\colhead{$K$} & \colhead{Frequency (GHz)} & \colhead{$g_{IK}$ \tablenotemark{a}} & 
\colhead{$S_{JK}$ \tablenotemark{b}} & \colhead{$g_{IK} S_{IK}$} & 
\colhead{$E_{JK}$ (K) \tablenotemark{c}} & \colhead{Einstein $A$ (s$^{-1}$)} 
}
\startdata
2 & 220.730266 &  6 & 140/12 & 70.0 & 97.4 & $8.98 \times 10^{-4}$ \\
3 & 220.709024 & 12 & 135/12 & 135.0 & 133.2 & $8.66 \times 10^{-4}$ \\
4 & 220.679297 &  6 & 128/12 & 64.0 & 183.1 & $8.21 \times 10^{-4}$ \\
5 & 220.641096 &  6 &  119/12 & 59.5 & 247.4 & $7.63 \times 10^{-4}$ \\
6 & 220.594438 & 12 & 108/12 & 108.0 & 325.9 & $6.92 \times 10^{-4}$ 
\enddata
\tablenotetext{a}{\mbox{} Statistical weight.}
\tablenotetext{b}{\mbox{} Line strength.}
\tablenotetext{c}{\mbox{} Upper level energy.}
\end{deluxetable}

\section{RADIATIVE TRANSFER MODELS \label{models}}
In order to derive the density and temperature profiles, spherically symmetric
models are developed for self-consistency to explain the continuum and methyl cyanide 
line emissions of each source A and C.
The density and temperature are assumed to have power-law dependences on radius,
\begin{eqnarray}
\rho &=& \rho_0 \, \left(\frac{r}{r_0}\right)^{-p}~\mathrm{g \, cm^{-3}} \\
n &=& n_0 \,  \left(\frac{r}{r_0}\right)^{-p}~\mathrm{cm^{-3}} \\
T &=& T_0 \, \left(\frac{r}{r_0}\right)^{-q}~\mathrm{K},
\end{eqnarray}
where $\rho_0$, $n_0$ and $T_0$ are the mass density of dust, number density 
of methyl cyanide molecules, and temperature at the reference radius 
$r_0 = 500 \; \mathrm{AU}$, respectively.
The gas and dust components are assumed to have the same temperature, 
$T_g = T_d = T$, everywhere in our models.
This assumption will be discussed in detail in Sect.~\ref{sec_coupling}.

Model images which are convolved with the observed angular resolutions 
are used to compare with the maps.
First, intensity images are made by integrating the intensity, $I_{\nu}$, along the 
line of sight (LOS) with the radiative transfer equation
\begin{equation}
\label{eq_rte}
\frac{dI_{\nu}}{d\tau_{\nu}}  = -I_{\nu} + S_{\nu} 
\end{equation}
where $S_{\nu}$ is the source function and $\tau_{\nu} = \int \alpha_{\nu} dz$ is 
the optical depth, which is an integral of the absorption coefficient $\alpha_{\nu}$ 
along the LOS.
The Boltzmann level population is assumed for the methyl cyanide lines (see the
discussion in Sect.~\ref{sec_tlp}). 
Hence the source function is the Planck function, $S_{\nu} = 2 k T \nu^2 / c^2$,
for both the continuum and methyl cyanide line models.
The intensity images are then convolved with the restored Gaussian beam to 
generate model images that can better simulate the observations.
In order to select the model that best describes the observation results, 
the chi-square, $\chi^2$, is calculated and optimized with the Levenberg-Marquardt 
method.
The goodness of the fits is evaluated by the reduced chi-square, 
$\bar{\chi^2} \equiv \chi^2/f$, where $f$ is the degrees of freedom in the fits.
Since the expectation value of the chi-square is equal to the degrees of 
freedom, $\chi^2 \simeq f$, it is easier to compare the results with the reduced chi-square,
whose expectation value is unity, $\bar{\chi^2} \simeq 1$, for a moderately good fit.

Since our methyl cyanide observations did not resolve each individual component, 
we have chosen two power-law indices, $q = 0.4$ and $0.75$, corresponding to 
the optically thin and thick regimes to the infrared (IR) photons, to investigate 
the density profiles and other physical quantities.
We will explain how we determine the range of $q$ in Sect.~\ref{Sec_tp}.
The dust distribution is assumed to be a thick shell-like structure with an 
inner radius $R_{\mathrm{cav}}$, which may correspond to the dust destruction radius, 
and an outer radius $R_d$, beyond which no dust is present.
Four parameters are fitted for the continuum models: $\rho_0$, $p$, $R_d$, 
and $R_{\mathrm{cav}}$, while $q$ and $T_0$ are held fixed.
The methyl cyanide line models are also fitted with four parameters: $n_0$, $T_0$, 
the Gaussian line width, $b$, and the radial extent of methyl cyanide, 
$R_{\mathrm{CH_3CN}}$.
Again, two parameters, $q$ and $R_{\mathrm{cav}}$, of line models are held 
unchanged during optimization.
The line profile, $\phi_{\nu}$, is assumed to be a Gaussian function with a half-width, $b$,
to $1/e$:
 \begin{equation}
 \label{eq_phi}
 \phi_{\nu} = \frac{1}{\sqrt{\pi} \, b} \, e^{-\left(\frac{\nu - \nu_0}{b}\right)^2}.
 \end{equation}
Lastly, the continuum and line models are iterated until self-consistency is obtained 
for the three parameters, $T_0$, $p$, and $R_{\mathrm{cav}}$ that are used in both models.

Considering that molecules can not survive in the stellar ultraviolet (UV) radiation without 
shielding by dust grains, the sizes of the central cavities, $R_{\mathrm{cav}}$, are 
assumed to be the same for the continuum and line models.
Nevertheless, the derived temperatures are insensitive to the sizes of 
$R_{\mathrm{cav}}$ due to the large beam size of 1\arcsec.
The values of $T_0$ do not fluctuate more than 5~K when $R_{\mathrm{cav}}$ 
increases from 20 to 300~AU.

\section{CONTINUUM OBSERVATION RESULTS \label{sec_cont}} 
\subsection{Grain growth in \wtw}
\begin{figure}
\plotone{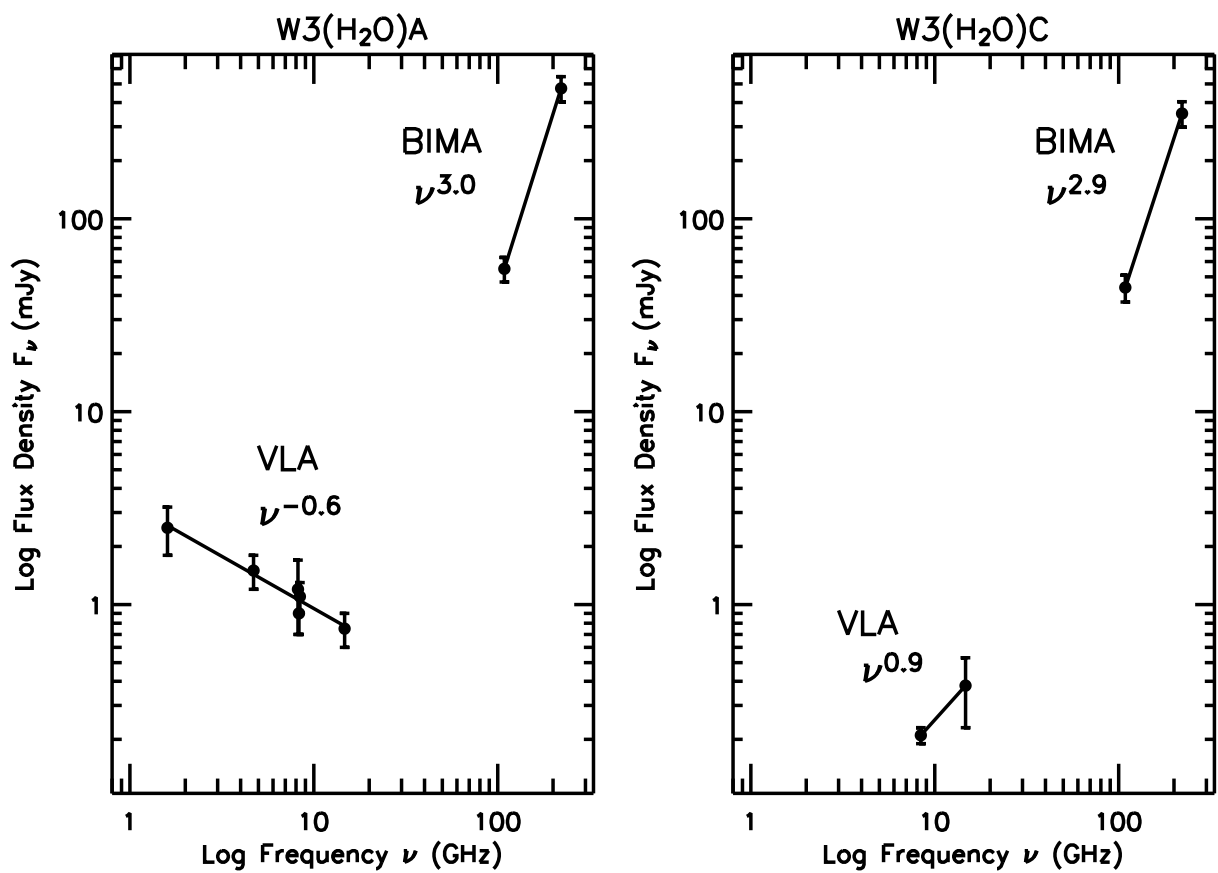}
\caption{Continuum spectra of the two components in W3(H$_2$O).   Dust thermal emission 
is detected at millimeter wavelengths.   Both components show rising spectra 
with spectral indices of $3.0 \pm 0.4$ and $2.9 \pm 0.4$, 
which are smaller than the spectral index of 3.6 from previous studies.  
At lower frequencies, the east component W3(H$_2$O)A shows synchrotron emission \markcite{Reid95}({Reid} {et~al.} 1995)
while the west component W3(H$_2$O)C has a rising spectrum indicating another possible 
protostellar object \markcite{Wilner99}({Wilner} {et~al.} 1999). \label{contsp} }
\end{figure}

The continuum spectra of the two sources from 1.6~GHz to 221.2~GHz are 
shown in Fig.~\ref{contsp}.
The flux densities at 1.4~mm and 2.8~mm are found from this study by integrating 
the flux in a circular region of 0\farcs5 radius centered at each peak.
The results are listed in Table~\ref{contflux}.
The 15\% systematic errors are included in the uncertainties.
The spectral indices are $3.0 \pm 0.4$ for source~A and $2.9 \pm 0.4$ for source~C.
In summary, both sources show a rising spectrum between 1.4~mm and 2.8~mm 
with a spectral index of approximately 3.0.
The spectral indices we observed here are smaller than, though consistent with,  
a spectral index of 3.6 in previous studies, which were based on a weak measurement 
at 3.4~mm \markcite{Wilner94}({Wilner} \& {Welch} 1994).
The improved sensitivity and angular resolution give us more reliable flux density 
measurements and the frequency dependence of the opacity can be determined for 
both sources. 

\begin{deluxetable}{ccccccc}
\tablewidth{0pt}
\tablecolumns{7}
\tablecaption{Millimeter Continuum Flux Densities and Spectra \label{contflux}}
\tablehead{ \colhead{} & \multicolumn{2}{c}{Coordinate (J2000)} & 
    \colhead{$F_{\mathrm{1.4~mm}}$} & \colhead{$F_{\mathrm{2.8~mm}}$} & \colhead{} \\
   \colhead{Source} & \colhead{RA} & \colhead{Dec} & \colhead{(mJy)} & 
   \colhead{(mJy)} & \colhead{$F_{\nu}$} }
\startdata
A & 2$^{\rm h}$27$^{\rm m}$4\fs709 & +61\arcdeg52\arcmin24\farcs64 & 
        $474 \pm 71$ & $55 \pm 8$ & $\nu^{3.0 \pm 0.4}$ \\
C & 2$^{\rm h}$27$^{\rm m}$4\fs541 & +61\arcdeg52\arcmin24\farcs70 & 
        $351 \pm 53$ & $44 \pm 7$ & $\nu^{2.9 \pm 0.4}$ \\
\enddata
\end{deluxetable}

If the dust thermal emission is optically thin, containing a factor $\nu^2$
from the Planck function, a millimeter spectrum of 
$F_{\nu} \propto \nu^{3.0 \pm 0.4}$ will give a mass opacity law of 
$\kappa_{\nu} \propto \nu^{\beta}$ with $\beta = 1.0 \pm 0.4$.
On the other hand, a smaller spectral index could be a result of dust emission 
being optically thick at 1.4~mm.
We will use our continuum models to show that the averaged optical depths 
are less than $\langle \tau \rangle \simeq 0.09$ in the cores (see Sect.~\ref{subsec_dp}
and Table~\ref{pflist}).
Hence, the small spectral indices are not the result of large optical depths.

The frequency dependence of the dust opacity $\kappa_{\nu}$ at long wavelength is 
an important parameter that can distinguish the particle types such as their shapes 
and mixtures.
Dust grains in various environments may differ dramatically.
The diffuse interstellar medium, consisting a mixture of small silicate and graphite spheroids, 
has a dust opacity spectral index of 2 at wavelengths longer than 100~$\mu$m \markcite{Draine84}({Draine} \& {Lee} 1984).
In circumstellar disks, random aggregation of small dust grains may result in the formation 
of large particles with fractal dimensions.
When grain growth happens, the emissivity at long wavelengths will be enhanced 
(relative to that at short wavelengths) and the spectral index of the opacity may 
decrease down to $\lesssim 1$ \markcite{Wright87,Miyake93}({Wright} 1987; {Miyake} \& {Nakagawa} 1993).
A smaller spectral index at millimeter wavelengths has been proposed as evidence of 
grain growth in dense cores and in the circumstellar disks around low-mass 
pre-main-sequence stars 
\markcite{Beckwith90,Chandler95,Beuther04}({Beckwith} {et~al.} 1990; {Chandler} {et~al.} 1995; {Beuther}, {Schilke}, \& {Wyrowski} 2004).
In a high density environment, collision and sticking can be much more effective to encourage 
grain growth.
The observed dust opacity law of  $\kappa_{\nu} \propto \nu^{1.0 \pm 0.4}$ suggests possible grain growth in the vicinities of the massive protobinary system in \wtw. 

\subsection{Choice of the Temperature Profiles \label{Sec_tp}}
Temperature profiles of dust cocoons around embedded massive stars remain 
one of the most difficult questions that need to be answered.
Since the emission distribution is an interplay between the temperature and density 
distributions, the temperature profile directly affects the extraction of the density
profile.
The UV radiation from the protostar is absorbed by the dust cocoon and 
re-radiated as dust thermal emission, whose spectrum peaks in the IR.
In general, the temperature profile reflects the diffusion of the IR photons.
The innermost part of the core is very dense and optically thick to the IR photons while 
the outer part of the core should be optically thin.
The transition between the two regimes possibly happens around \mbox{$r \sim 10^3$ AU}.
To date, observations have been limited by angular resolution and have mainly focused 
on the density profiles in the outer regime, where the temperature profile has a simple radial 
dependence of \mbox{$r^{-0.4}$}.
The temperature of the inner regime can only be calculated with radiative transfer models.
However, simple theoretical calculations do not account for a wide variety of processes 
that may occur in accreting flows around massive protostars, for example, 
radial variation of dust properties, effects of magnetic field, etc.   
Besides, the feedback of numerous chemical reactions to the heating and cooling processes 
has generally been neglected.  
Our dust continuum observations have partially resolved the inner region, where
the temperature profile may differ significantly from that of the outer regime.
We have chosen two extreme cases, optically thin and optically thick, to investigate 
the corresponding parameter spaces. 
 
In the outer regime, theoretical studies \markcite{Wolfire86, Osorio99}({Wolfire} \& {Cassinelli} 1986; {Osorio}, {Lizano}, \& {D'Alessio} 1999) have shown 
that the temperature profile is expected to asymptotically approach a power-law 
index of  $q \simeq 0.4$ and is a weak function of the dust opacity law 
\markcite{Wolfire86}({Wolfire} \& {Cassinelli} 1986).
Assuming that the dust thermal emission can be characterized by a modified Planck
function of a single temperature $T_D$, we can write the local radiative 
equilibrium equation as
\begin{equation}
\int_0^{\infty} \kappa_{\nu}B_{\nu}(T_D) d\nu = \frac{\mathrm{const}}{r^2}.
\end{equation}
If the dust opacity law takes the form of $\kappa_{\nu} \propto \nu^{\beta}$, the integration 
gives a temperature distribution of $T_D \propto r^{-2/(4+\beta)}$.
Since surveys in star-forming regions usually measure the opacity law to be in the range of
$\beta = 1 - 2$, the power-law index of the temperature profiles therefore falls into
the range of  $q = 0.4 - 0.33$. 
The observed overall dust opacity law of $\kappa_{\nu} \propto \nu$ in \wtw\ will have 
$q = 0.4$ in the optically thin regime.

In the inner regime where the optical depth is large for the IR photons,
one can derive a simple relationship between the density and temperature distributions 
using the standard diffusion approximation 
\markcite{Osorio99,Kenyon93a}({Osorio} {et~al.} 1999; {Kenyon}, {Calvet}, \& {Hartmann} 1993)
\begin{equation}
\label{diffusion}
L = -\frac{64 \pi \sigma_{SB} \, r^2 T^3}{3 \kappa_R \rho} \frac{dT}{dr} = \mathrm{const.}
\end{equation}
where $L$, $\rho$, $\sigma_{SB}$ are the luminosity, density, and Stefan-Boltzmann constant, respectively.
The Rosseland mean opacity $\kappa_R(T)$ is an averaged opacity defined as
\begin{equation}
\frac{1}{\kappa_R} \equiv \frac{\displaystyle \int_0^{\infty} \frac{1}{\kappa_{\nu}} 
                                            \frac{\partial B_{\nu}}{\partial T} d\nu}
                                   {\displaystyle \int_0^{\infty} \frac{\partial B_{\nu}}{\partial T} d\nu }.
\end{equation}
When the opacity law is $\kappa_{\nu} \propto \nu$, the Rosseland mean opacity can be easily
calculated to be $\kappa_R(T) \propto T$.
Assuming a power-law dependence on radius for the density and temperature distributions, 
$\rho \propto r^{-p}$ and $T \propto r^{-q}$, one can apply the Rosseland mean
opacity of $\kappa_R(T) \propto T$ to Eq.~(\ref{diffusion}) and obtain a temperature profile 
with a power-law index of $q = (p+1)/3$.
Given the observed radial emission profile with a power-law index of approximately 
$2 = p+q$ (see Sect.~\ref{subsec_dp}), the largest possible value for $q$ will be 0.75.
Although the exact temperature behavior is more complicated than merely a single 
power-law, we have chosen the two extreme cases, $q=0.4$ and $0.75$, for 
our interpretation of the data. 

\subsection{Emissivity Profiles}
\begin{figure}
\plotone{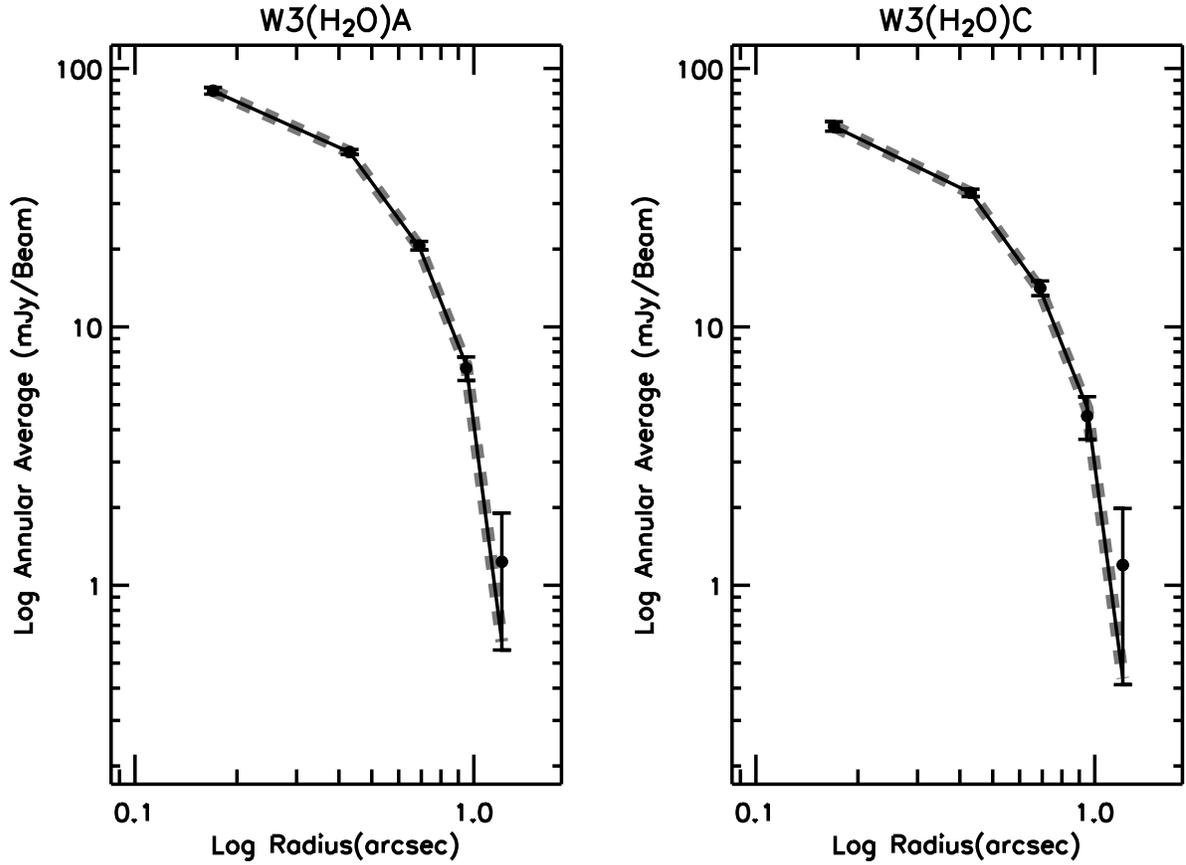}
\caption{The 1.4~mm continuum emission profiles of annular averaged intensities versus
the outer radii of the annuli.
The annular averages are taken with a spacing of the full synthesized beam 
so the data points are largely independent.
The optimized solution for $q = 0.4$ (solid curve) and that for $q = 0.75$ 
(dashed curve) fit the data equally well.
The cutoffs at large radii are caused by the finite sizes of the dust distributions.
The turnovers at the innermost radii result from a combined effect of 
the central cavities and the convolution with the beam.
\label{ringq}}
\end{figure}

Since the 1.4~mm beam (0\farcs26) is about 5 times smaller than the extent of each source 
in \wtw, continuum emission profiles can be calculated by taking averaged 
intensities in successive annuli centered at each continuum peak.
Fig.~\ref{ringq} shows the emission profiles by plotting the averaged intensities against 
the outer radii of the annuli with a spacing of one synthesized beam to avoid
undesired correlation between data points.
In order to reduce the interference between the two sources, we have masked one source 
while taking annular averages on the other source.
The irregular extended emission to the northwest is also excluded from the emission profiles of 
source~C.
The uncertainty of each annular average, $\sigma_i$, is defined as the ratio of the 
root-mean-square (rms) in the map to the square root of the number of beams within 
the annulus of interest: 
\begin{equation}
\sigma_i \equiv \frac{\sigma}{\displaystyle \sqrt{\frac{N_i}{N_{\mathrm{Beam}}} } },
\end{equation}
where $\sigma$ is the rms of the map; $N_i$ and $N_{\mathrm{Beam}}$ are the numbers 
of pixels in the $i$th annulus and one beam, respectively.

\subsection{Density Profiles \label{subsec_dp}}
The optimized solutions are listed in Table~\ref{pflist} for the two selected $q$ values and 
the results are shown in Fig.~\ref{ringq}.
Overall, the emission profiles have power-law indices of approximately -2 for both sources.
The emissivity is proportional to the product of the temperature and density as 
$\epsilon(r) \propto \rho_0 \, T_0 \, r^{-(p+q)}$.
If the temperature distribution follows $T \propto r^{-0.4}$, the underlying density profile will 
have a power-law index of $p \simeq 1.6$. 
The density distributions may have an even smaller index of $p \simeq 1.2$ if the 
temperature profile is as steep as $T \propto r^{-0.75}$.
In order to fit the innermost data point at 0\farcs17, we introduce a dust-free cavity of radius 
$R_{\mathrm{cav}}$, which may correspond to the dust destruction radius.
The value of $R_{\mathrm{cav}}$ is in the range of $200$ to $250$~AU, much smaller than the angular resolution 
of 0\farcs26 ($\simeq 570$~AU).
The real temperature distribution shall deviate greatly from a single power-law
temperature profile, especially near the dust destruction radius.
A reliable estimate of $R_{\mathrm{cav}}$ will need better resolution and more 
elaborate calculations.   
\begin{deluxetable}{lcccc}
\tablewidth{0pt}
\tablecolumns{5}
\tablecaption{Optimized $\bar{\chi^2}$ Fits for the Continuum Models \label{pflist}}
\tablehead{ 
\colhead{Parameter} & \multicolumn{2}{c}{$q = 0.4$} & \multicolumn{2}{c}{$q = 0.75$}  \\
\colhead{Source} & \colhead{A} & \colhead{C} & \colhead{A} & \colhead{C} 
}
\startdata
$\bar{\chi^2}_{min}$ & 1.00 & 1.14 & 1.00 & 1.15 \\
$p$ & 1.52 $\pm$ 0.17  & 1.52 $\pm$ 0.27 & 1.15 $\pm$ 0.16 & 1.14 $\pm$ 0.27 \\
$\rho_0$ ($10^{-18} \; \mathrm{g \, cm^{-3}}$)  & 
3.59 $\pm$ 0.42 & 2.78 $\pm$ 0.47 & 3.48 $\pm$ 0.36 & 2.48 $\pm$ 0.41 \\
$R_d$ ($10^3$ AU) & 2.05 $\pm$ 0.09 & 2.06 $\pm$ 0.16 & 2.05 $\pm$ 0.09 & 2.06 $\pm$ 0.16 \\
$R_{\mathrm{cav}}$ ($10^2$ AU) & 2.48 $\pm$ 0.51 & 2.17 $\pm$ 0.83 & 2.50 $\pm$ 0.46 & 2.18 $\pm$ 0.81 \\
\sidehead{Parameters set by the $\mathrm{CH_3CN}$ models}
$T_0$ (K) & 197 & 172 & 205 & 192 \\ 
\tableline\tableline
\sidehead{Parameters derived from models:}
$\langle \tau \rangle \tablenotemark{a}$ & 0.094 & 0.075 & 0.092 & 0.071 \\
$M_d$ ($\mathrm{M_{\sun}}$) & 0.043 & 0.034 & 0.057 & 0.041 \\
$M_g$ ($\mathrm{M_{\sun}}$) \tablenotemark{b} & 4.3 & 3.4 & 5.7 & 4.1 \\
$n_{\mathrm{H_2}}$ ($10^8 \; \mathrm{cm^{-3}}$) \tablenotemark{b}\tablenotemark{c}
              & 1.07 & 0.83 & 1.04 & 0.74 \\
\enddata
\tablenotetext{a}{Averaged optical depth, $\langle \tau \rangle$, 
within 0\farcs5 around the peak.}
\tablenotetext{b}{A gas-to-dust ratio of $100$ is assumed.}
\tablenotetext{c}{The number densities of H$_2$ at $r_0$, calculated from $\rho_0$.}
\end{deluxetable}

The self-similar collapse of an isothermal sphere (SIS) \markcite{Shu77}({Shu} 1977) is able to explain the 
collapse of low-mass protostellar cores.
This inside-out model describes the quiescent part of an isothermal cloud core with 
a density profile of $\rho \propto r^{-2}$ and propagates the collapse signal with 
an expansion wave, behind which a $\rho \propto r^{-3/2}$ law holds for the 
freely falling inner envelope. 
On the other hand, earlier observations of massive star-forming cores on
larger scales suggested the mean density index $p$ in a range between 1.6 to 1.8 
with large scatters in the samples \markcite{vanderTak00a,Beuther02c,Mueller02}({van der Tak} {et~al.} 2000; {Beuther} {et~al.} 2002; {Mueller} {et~al.} 2002).
The cores may be supported by turbulence and magnetic fields \markcite{Myers92,McKee03}({Myers} \& {Fuller} 1992; {McKee} \& {Tan} 2003).
Studies of dense cores associated with massive stars, L1688 and L1204, \markcite{Myers92}({Myers} \& {Fuller} 1992) 
suggested that the support from turbulence or magnetic fields most likely dissipates 
on the scale of $0.01$~pc, where we observed the density profiles of the two sources 
in \wtw.
The small spatial extent of our density profiles may favor the explanation of free-fall collapse. 
Therefore, we suggest that both sources in \wtw\ are in an active phase of accreting material 
from their surroundings.

In order to estimate the optical depth of each clump, we calculate an averaged optical depth,
$\langle \tau \rangle$, which is weighted by the solid angle of each annulus,
\begin{equation}
       1 - e^{\langle \tau \rangle} \equiv 
        \frac{\int (1-e^{-\tau}) \, 2 \pi a \, da}{\int 2 \pi a \, da},
\end{equation}
where $a$ is the angular distance from a continuum peak.
The integral is performed within 0\farcs5 around each peak, the same region where 
we integrate the continuum flux density.
In general, the averaged optical depths are in the range of $0.07$ to $0.09$.
Considering a clump of uniform density and temperature with $\langle \tau \rangle = 0.09$, 
the error of the 1.4~mm flux density made by the optically thin approximation is
$\langle \tau \rangle/(1-e^{-\langle \tau \rangle}) \simeq 5\%$.
This in turn contributes an additional error of only $0.07$ in the spectral index.  
Therefore, the optically thin assumption should be reasonable and the small
spectral indices are not the result of large optical depths at 1.4~mm. 

\subsection{Mass Estimates}
Table~\ref{pflist} also lists the estimated masses and maximum optical depths 
of the dusty nebula around each source.
The mass of the dust, $M_d$, can be computed simply by integrating 
the density over the volume of the clump.
Once the mass of the dust is known, the clump (gas) mass, $M_g$, 
can be obtained by assuming a conventional gas-to-dust ratio of $M_g/M_d = 100$.
Over all, the nebular mass of source~A is in the range of 4 to 6\Msun\ while that of 
source~C is in the range of 3 to 4\Msun.
Nevertheless, these masses exclude any mass contained in the central compact objects.
The number densities of H$_2$, $n_{\mathrm{H_2}}$, at the reference radius, 
$r_0 = 500 \; \mathrm{AU}$, are also listed in Table~\ref{pflist}.
In general, the number densities are roughly $10^8 \; \mathrm{cm^{-3}}$, which 
are much higher than $n_{\mathrm{H_2}} \simeq 2 \times 10^6 \; \mathrm{cm^{-3}}$ as
suggested by \markcite{Helmich97}{Helmich} \& {van Dishoeck} (1997) in a study of single-dish spectral line surveys.

A gas-to-dust ratio of $M_g/M_d \simeq 100$ \markcite{Savage79}({Savage} \& {Mathis} 1979) 
is the most commonly accepted value, where it was derived from comparing 
the visual extinction with the hydrogen column density, and should be good within a 
factor of 2 \markcite{Hildebrand83}({Hildebrand} 1983).
This ratio serves as a coefficient to convert dust mass to gas (clump) mass and is
used in the mass estimates of Table 3.
The largest uncertainty in the density (or mass) determination comes from 
the chosen dust opacity, where various estimates of $\kappa_{\mathrm{1.4\,mm}}$ 
span a range from 0.2 to \mbox{$3 \; \mathrm{cm^2 \, g^{-1}}$} \markcite{Henning95,Beckwith91}({Henning}, {Michel}, \&  {Stognienko} 1995; {Beckwith} \& {Sargent} 1991).
Observational determinations of mass opacity require a few assumptions, each of which brings uncertainties into the final result \markcite{Hildebrand83}({Hildebrand} 1983).
In addition, various environments may carry dust grains of different types, which also 
contribute to the variation in the derived mass opacity.
On the other hand, theoretical calculations suffer from unknown dust shape and 
chemical composition.  
A mixture of small graphite and silicon spheroids may have a mass opacity as low as 
\mbox{$\kappa_{\mathrm{1.4\,mm}} \simeq 0.2 \; \mathrm{cm^2 \, g^{-1}}$} 
\markcite{Draine84}({Draine} \& {Lee} 1984) while grain growth with fractal dimension can 
increase $\kappa_{\mathrm{1.4\,mm}}$ by a factor up to 30 
\markcite{Wright87}({Wright} 1987).
Based on a more sophisticated model of the coagulation process 
\markcite{Ossenkopf93} ({Ossenkopf} 1993), \markcite{Ossenkopf94}
{Ossenkopf} \& {Henning} (1994) systematically computed and tabulated dust opacities 
in dense protostellar cores.
They suggested a dust opacity of $1.11 \; \mathrm{cm^2 \, g^{-1}}$ for a core of 
$n_{\mathrm{H_2}} = 10^8 \; \mathrm{cm^{-3}}$ with non-negligible warming by
near-IR radiation.

The opacity of \mbox{$\kappa_{\mathrm{1.4\,mm}} = 1.84 \; \mathrm{cm^2 \, g^{-1}}$} 
in our continuum models is calculated by applying the observed power-law indices of 
\mbox{$\beta \simeq 1.0$} while maintaining the normalization at \mbox{$250 \; \mu$m}:
\begin{equation}
\label{opacity}
\kappa_{\nu} = 10 \; \mathrm{cm^{2} \, g^{-1}}~
                         \left( \frac{\lambda_{\mathrm{mm}}}{0.25} \right)^{-\beta},
\end{equation} 
where the value of \mbox{$10 \; \mathrm{cm^{2} \, g^{-1}}$} at $250 \; \micron$ 
is given by \markcite{Hildebrand83}{Hildebrand} (1983).
This is a similar approach to what has been suggested by \markcite{Beckwith91}{Beckwith} \& {Sargent} (1991) for deriving the masses of circumstellar disks around low-mass young stars.
The dust opacity of $1.84 \; \mathrm{cm^2 \, g^{-1}}$ is a factor of 1.66 higher 
than the computed value of $1.11 \; \mathrm{cm^2 \, g^{-1}}$ 
(\markcite{Ossenkopf94}{Ossenkopf} \& {Henning} 1994) and may underestimate 
the nebular masses by 40\% if one adopts the computed opacity.

\section{METHYL CYANIDE LINE OBSERVATION RESULTS \label{sec_line}}
A temperature measurement is important to estimate the luminosity of an embedded 
protostellar object as well as to break the degeneracy of $\rho_0 \, T_0$ in our 
continuum models.
Temperature can be derived when multiple transitions of a molecular species 
with a great range of excitations are observed.
Symmetric-top molecules with large dipole moments, such as methyl cyanide 
($\mathrm{CH_3CN}$), are ideal ``thermometers" to probe the gas temperature.
This type of molecule has groups of lines with greatly different excitation energies 
but close in frequency, so that the spectral correlators can be adjusted to observe 
these transitions simultaneously.
This reduces the uncertainty from calibration, antenna gains, $uv$-sampling, etc. 
In addition, the radial velocity can be determined with high precision by measuring 
the Doppler shifts in all the transitions. 
Line observations with good spectral resolution are necessary to obtain kinematical clues for  understanding the relationship of the two continuum sources.

\subsection{Kinematics of the binary \label{sec_vel}}
\begin{figure}
\plotone{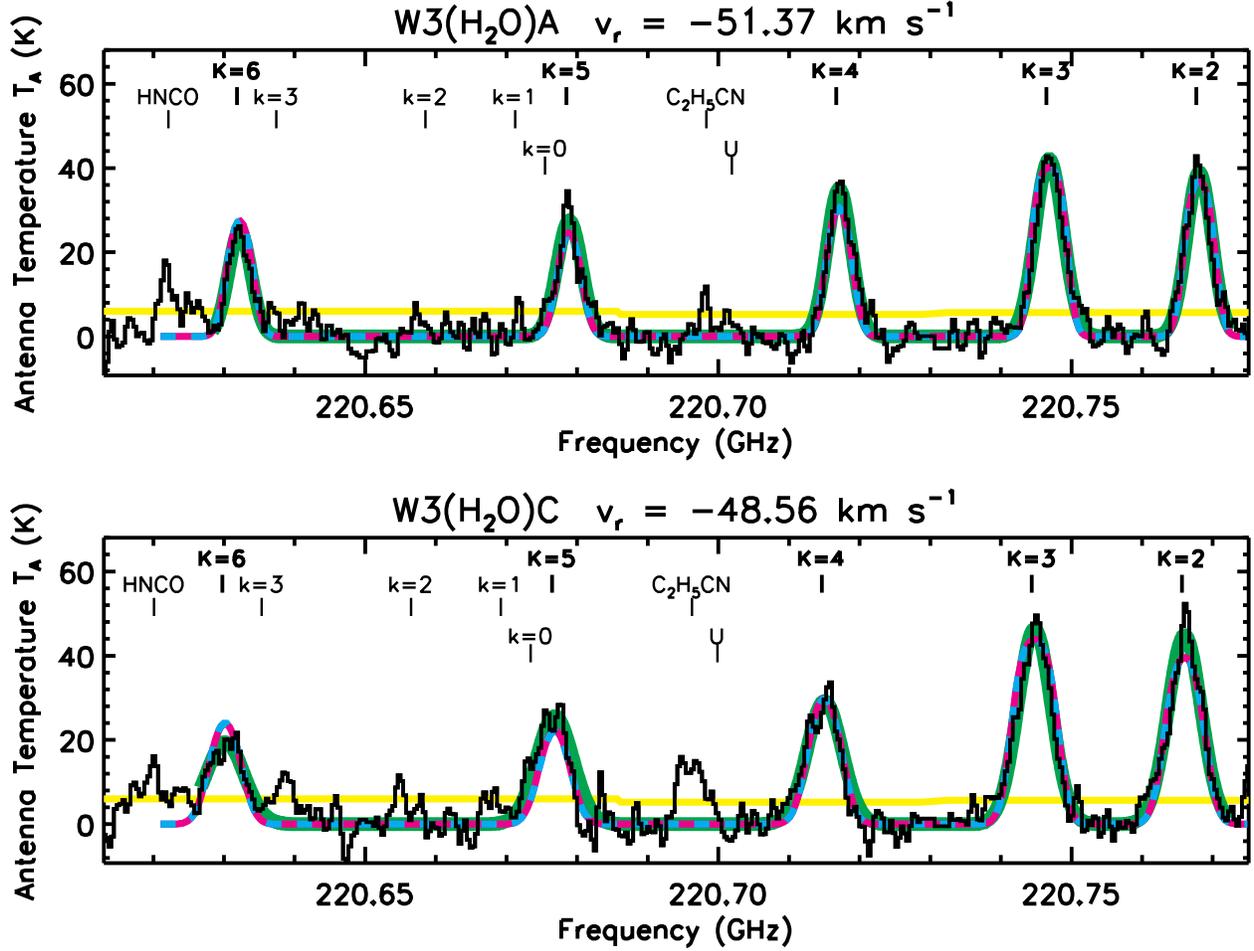}
\caption{Spectra of methyl cyanide emission observed at source A and source B 
(shown in histograms).
Only data points above the $2 \sigma$ level (yellow curves) are included for 
further analysis.
The radial velocities, $\upsilon_r$, are found by simultaneously fitting 5 Gaussians 
(green curves); each Gaussian has its own width and amplitude but the five $K$ components 
have the same radial velocity.
The optimized solutions for $q = 0.4$ are shown in blue dashed curves 
while those for $q=0.75$ are shown in magenta dashed curves.
The big $K$ values indicate the $K$ components of the 
$\mathrm{CH_3CN} \;\; J = 12 \rightarrow 11$ transitions and the small k values 
indicate the corresponding 
$K$ components of its isotopomer $\mathrm{CH_3\,^{13}CN}$. \label{mcqv} }
\end{figure}

Assuming that all the $K$ components have the same velocity, we can simultaneously 
fit Gaussians to each of the five $K$ components to trace the bulk motion 
of the source.
The results of our Gaussian fits are shown as the green curves in Fig.~\ref{mcqv}.
Only data above the $2\sigma$ level (yellow curves) are included in the fits. 
The velocities found in Gaussian fits are 
\mbox{$\upsilon_r = -51.37 \pm 0.05 \; \mathrm{km \, s^{-1}}$} at source~A and 
\mbox{$\upsilon_r = -48.56 \pm 0.05 \; \mathrm{km \, s^{-1}}$} at source~C.
Therefore, the velocity difference between the two continuum peaks is 
\mbox{$\Delta \upsilon_{obs} = 2.81 \pm 0.10 \; \mathrm{km \, s^{-1}}$}.

Consider binary motion consisting of two circular orbits at an inclination $i$.
The separation, $a$, of the binary members is the sum of the semi-major 
axes of the two orbits. 
Let the relative velocity between the binary be $\Delta \upsilon = \upsilon_1 - \upsilon_2$.
At the maximum projected separation, $a$, the radial velocity difference is also the largest,  
$\Delta \upsilon_r = \Delta \upsilon \, \sin i$.
The total mass of the binary can be expressed as 
\begin{equation}
\label{Mbinary}
M_{\mathrm{binary}} = M_1 + M_2 = \frac{(\Delta \upsilon_r)^2 \, a}{G} \frac{1}{\sin^2 i},
\end{equation}
where \mbox{$G=6.673 \times 10^{-8} \; \mathrm{g^{-1} \, cm^3 \, s^{-2}}$} is the 
gravitational constant.
Suppose that the two sources in the hot core happened to be at their maximum projected 
separation. 
The binary separation is just the observed separation, 
\mbox{$a = a_{obs} = 2.43 \times 10^3 \; \mathrm{AU}$}, and the radial velocity difference is 
\mbox{$\Delta \upsilon_r = \Delta \upsilon_{obs} = 2.81 \pm 0.10 \; \mathrm{km \, s^{-1}}$}.
Since the trigonometric function in Eq.~(\ref{Mbinary}) is always less than unity, 
the minimum binary mass will be
\begin{eqnarray}
M_{\mathrm{W3(H_2O)}} \ge \frac{(\Delta \upsilon_r)^2 a}{G} 
     &=&1.13 \; \left( \frac{\Delta \upsilon_{obs}}{\mathrm{km \, s^{-1}}} \right)^2 
        \left( \frac{a_{obs}}{\mathrm{10^3 \; AU}} \right)~\Msun \nonumber \\
     &=& 22\Msun.
\end{eqnarray}
This sets a lower limit of 22\Msun\ for the total mass of the two protostars.
The radial velocity difference that we observed here is in a reasonable range 
for binary motion.
Therefore, we suggest that the two continuum peaks are members of a binary and the 
velocity difference is due to the binaries orbiting around each other.

\markcite{Sutton04}{Sutton} {et~al.} (2004) have also observed a small radial velocity difference of 
$1.6 \; \mathrm{km \, s^{-1}}$ between source~A and C with several methanol lines 
at low spatial resolution.
They suggested that the velocity difference is caused by binary motion.
The smaller velocity difference should be a result of averaging over a larger solid angle.
Besides, methanol emission seems to emerge from a region larger than that of  
methyl cyanide. 
The geometry and kinematical structures of the circumstellar envelopes 
will also affect the observed radial velocities.
 
\subsection{Boltzmann Population of Levels \label{sec_tlp}}
The methyl cyanide spectra (Fig.~\ref{mcqv}) clearly show a suppression in the $K=3$ 
component. 
The ratio of the peak antenna temperature of $K=2$ to that of $K=3$ is rougly unity, 
which is much smaller than the expected optically thin
ratio of roughly 2 based on the product of the statistical weight and line strength, 
$g_{IK} S_{IK}$ (Table~\ref{obsK}).
It is evident that the $K=3$ component is optically thick and its antenna temperature 
should be close to the kinetic temperature of the gas if the emission fills 
the beam.
On the other hand, the $K=6$ component with an upper energy level of 
$E_{\mathrm{up}} = 326$~K is also detected at both sources, suggesting the 
presence of hot gas components.
Nonetheless, the observed antenna temperature of the $K=3$ line 
is only $T_A \simeq 60$~K, 
which is insufficient to explain the detection of the $K=6$ component. 
Therefore, the methyl cyanide emission must come from a region significantly smaller than 
our 1\arcsec~beam, equivalent to a radius of $10^3 \; \mathrm{AU}$.

Although the emission is spatially unresolved, the antenna temperature of the optically 
thick lines can still be used as a lower limit of the kinetic temperature of H$_2$.
Given this minimum value of 60~K, we can show that the observed 
methyl cyanide lines satisfy the conditions for LTE, which will give the Boltzmann level
population without further assumptions.
The collision rate coefficients, $K_{ul}$, of the observed lines are only involved with 
the fundamental rates $Q(L,M=0)$ (see Sect.~\ref{sec_pmc}).
In the case of $\mathrm{CH_3CN}$ $(12,K) \rightarrow (11,K)$ transitions, $K_{ul}$ can 
be estimated by its largest leading term in Eq.~(\ref{Kul_n0})
\begin{equation}
K_{ul} \, \gtrsim \, 23 \left( \begin{array}{ccc} 11 & 1 & 12 \\ K & 0 & -K \end{array} \right)^2 Q(1,0), 
\end{equation}
where the $3-j$ symbol is about 1.5 for all the $K$ components.
At a temperature of $60$~K, $Q(1,0) = 3.22 \times 10^{-10} \; \mathrm{cm^3 \, s^{-1}}$ 
so the collision rate coefficients will be 
$K_{ul} \gtrsim 1.67 \times 10^{-8} \; \mathrm{cm^3 \, s^{-1}}$. 
\markcite{Helmich97}{Helmich} \& {van Dishoeck} (1997) estimated a molecular hydrogen density of 
$n_{\mathrm{H_2}} = 2 \times 10^6 \; \mathrm{cm^{-3}}$ towards \wtw.
Therefore, the minimum value for downward collision rate will be 
$C_{ul} \gtrsim 3.4 \times 10^{-2} \; \mathrm{s^{-1}}$.  
This minimum rate of $C_{ul}$ will be even faster if the number densities 
of $n_{\mathrm{H_2}} \simeq 10^{8} \; \mathrm{cm^{-3}}$ in Table~\ref{pflist} are used.
On the other hand, the maximum value of the spontaneous emission rates among 
the observed lines is $A_{ul} = 9 \times 10^{-4}\; \mathrm{s^{-1}}$ from the $K=2$ 
component (Table~\ref{obsK}).
In other words, the collisional de-excitation rates are faster than the 
spontaneous emission rates by at least a factor of 40.
The transitions that we consider here should have level populations that follow 
the Boltzmann distribution at the kinetic temperature of H$_2$. 
 
\subsection{Model Results}
The velocity of each source has been obtained in the kinematics study (Sect.~\ref{sec_vel}).
We have assumed that methyl cyanide density has the same power-law dependence 
on radius as that of the dust. 
However, the spatial extent of the methyl cyanide emission, $R_{\mathrm{CH_3CN}}$, 
is smaller than that of the continuum emission, $R_{\mathrm{cont}}$.
Since the extent of the methyl cyanide is unknown, the emission size, 
$R_{\mathrm{CH_3CN}}$, is a parameter determined by the chi-square 
($\chi^2$) minimization.
The optimized solutions for the two selected $q$ values are listed in Table~\ref{mclist} 
and the results are shown in Fig.~\ref{mcqv}.
The $\bar{\chi^2}$ values for source~A are smaller than those for source~C.
In general, the values of $q$ do not dramatically change the goodness of the fits. 

Over all, the Gaussian line width $b$ is much larger than what one expects 
in the case of merely thermal motion, 
$b_{th} = (k T/m_{\mathrm{CH_3CN}})^{1/2} \simeq 0.07 \; \mathrm{km \, s^{-1}}$ 
at $T = 200$~K.
A velocity dispersion greater than the thermal velocity dispersion has been 
observed in low-mass dense cores within the scale at which turbulence dissipation 
takes place \markcite{Goodman98}({Goodman} {et~al.} 1998).
However, the thermal component of the line width comprises only 
an insignificant fraction of the total line width in our case.
It is not clear what mechanism is responsible for the nonthermal component: 
outflow, rotation, infall collapse, or small-scale turbulence may all contribute 
to the line width.
Besides, the velocity gradient between source A and C or the overlap of the two clumps
may also increase the line widths. 

The line width at source~C ($3.5 \; \mathrm{km \, s^{-1}}$) is much larger than 
that at source~A ($2.8 \; \mathrm{km \, s^{-1}}$), suggesting an active, embedded source.
Chemical differentiation has also been observed: 
\markcite{Wyrowski99}{Wyrowski} {et~al.} (1999) have detected the $v_7 = 2$, $J=24-23$ transition of
$\mathrm{HC_3N}$ (excitation at 722~K above the ground) only at source~C and the  $\mathrm{C_2H_5CN}$ ($24_{2,22} - 23_{2,21}$) emission peaked at source~C.
Our spectra also show a stronger $\mathrm{C_2H_5CN}$ ($25_{2,24} - 24_{2,23}$) 
emission at source~C.
This additional evidence also supports source~C being a more active source 
rather than just an over-dense clump.

The $\mathrm{CH_3\,^{13}CN}$ lines can be used to constrain the optical depths 
of the main isotopomer lines.
Comparing the measured antenna temperatures of the $K=2$ components of the 
two isotopomers, the $\mathrm{CH_3CN}$ line is brighter than the 
$\mathrm{CH_3\,^{13}CN}$ line by roughly a factor of 4.4.
The intensity ratio between the two species can be approximated by
$(\mathrm{CH_3\,^{13}CN}/\mathrm{CH_3CN}) = 
(\mathrm{^{13}C}/\mathrm{^{12}C}) \, [\tau/(1-e^{-\tau})]$,
where $\tau$ is the optical depth of the $\mathrm{CH_3CN}$ line.
Assuming a distance of $D_{\mathrm{GC}} = 10 \; \mathrm{kpc}$ to the Galactic Center,
the carbon isotope ratio is about $(\mathrm{^{12}CO}/\mathrm{^{13}CO}) = 82.6$ 
in the vicinity of W3(OH) (\markcite{Wilson94}{Wilson} \& {Rood} 1994).
Therefore, the optical depth of the $K=2$ component of the $\mathrm{CH_3CN}$ line 
should be less than $\tau \lesssim 19$.

\subsection{Mass Estimates}
The virial mass, $M_{\mathrm{vir}}$, is a function of the line width, $b$, the power-law index, $p$, 
of the density distribution, and the size of methyl cyanide emission, $R_{\mathrm{CH_3CN}}$
(see Appendix~\ref{sec_Mvir}):
\begin{eqnarray}
M_{\mathrm{vir}} &=& 
    \frac{3}{2} \left( \frac{5-2p}{3-p} \right) \left( \frac{b^2 \, R_{\mathrm{CH_3CN}} } {G} \right) \\
         &=& 0.17 \left( \frac{5-2p}{3-p} \right) \left( \frac{b}{\mathrm{km \, s^{-1}}} \right)^2 \left( \frac{R_{\mathrm{CH_3CN}}}{\mathrm{100 \; AU}} \right) \Msun.
\end{eqnarray}
On average, the virial mass of source A is about 12\Msun\ while the mass of 
source~C is about 28\Msun\ (Table~\ref{mclist}).
The derived virial masses are large due to the large line widths and should be 
considered as upper limits for the mass.
A massive star-forming core is an entity of complicated kinematics so a dynamical 
equilibrium most likely has not yet been achieved considering the youth of a hot core. 
The virial theorem is based on two fundamental assumptions: first, the objects form a relaxed 
system whose motion is gravitationally bound; second, the lines are not significantly 
broadened by other effects, such as large optical depth, systematic velocity gradients, etc.
Any deviation from the above two conditions will increase the value of the 
derived virial mass.

\subsection{Luminosity Estimates \label{sec_Ls}}
The simplest approach for calculating the luminosity is to model the emission as
a black body like that from a stellar surface:
\begin{eqnarray}
\label{eq_L}
L &=& 4 \pi r_0^2 \, \sigma_{SB} T_0^4  \nonumber \\
   &=& 1.0\times 10^3 \, \left( \frac{r_0}{\mathrm{500~AU}} \right)^2 
             \left( \frac{T_0}{\mathrm{100~K}} \right)^4 \Lsun,  
\end{eqnarray}
where $\sigma_{SB} = 5.67 \times 10^{-5} \; \mathrm{erg \, cm^{-2} \, s^{-1} \, K^{-4}}$ 
is the Stefan-Boltzmann constant.
This will give only a rough estimate of the luminosity.  
The luminosity for source~A is in the range of $(1.5-1.8) \times 10^4$\Lsun\ while
that of source~C is in the range of $(0.9-1.4) \times 10^4$\Lsun, 
equivalent to zero-age-main sequence (ZAMS) spectral type B0.5 - B0 stars 
\markcite{Panagia73}({Panagia} 1973).
Previous studies have also suggested a total luminosity of $10^4$\Lsun\ for the 
entire hot core \markcite{Turner84,Wilner95,Wyrowski99}({Turner} \& {Welch} 1984; {Wilner} {et~al.} 1995; {Wyrowski} {et~al.} 1999).
The stellar masses of ZAMS spectral type B0.5 - B0 stars are in the range of 
$11$ to $15 \; \mathrm{M_{\sun}}$ \markcite{hanson97}({Hanson}, {Howarth}, \& {Conti} 1997).
Therefore, the total mass of the binary system is in the range of 
$22$ to $30 \; \mathrm{M_{\sun}}$, in good agreement with the minimum 
kinematical mass of 22\Msun\ (Sec.~\ref{sec_vel}). 

An interesting question that arises with high luminosity is whether the radiation pressure 
acting on dust grains is large enough to halt the infall \markcite{Larson71,Kahn74}({Larson} \& {Starrfield} 1971; {Kahn} 1974).
Within factors of order unity, radiation pressure can reverse spherical infall if the luminosity 
to mass ratio, $L/M$, of the protostar exceeds a critical value of 
$700 \; \mathrm{L_{\sun}/M_{\sun}}$.
This critical value, however, varies with the assumed dust properties and 
the geometry of infall processes (see, {\it e.g.}, the review by \markcite{Shu87R}{Shu}, {Adams}, \& {Lizano} (1987)).
Over all, the ratios of luminosity to nebular mass, $L/M_g$, 
in our model results (Table~\ref{mclist}) are larger than this critical value but 
within a factor of 5.
The masses contained in the central protostars also add uncertainties to the observed 
ratios.

When the expansion of an \HII\ region is considered, the opacity is mainly due to the 
scattering of free electrons, and the Eddington limit gives a luminosity to mass ratio of 
$L/M = 3.2 \times 10^4 \; \mathrm{L_{\sun}/M_{\sun}}$. 
Our luminosity to nebular mass ratios, $L/M_g$, are smaller than the Eddington limit 
by roughly one order of magnitude.

\begin{deluxetable}{lcccc}
\tablewidth{0pt}
\tablecolumns{5}
\tablecaption{Optimized $\bar{\chi}^2$ Fits for the Methyl Cyanide Line Models \label{mclist}}
\tablehead{ 
\colhead{Parameter} & \multicolumn{2}{c}{$q=0.4$} & \multicolumn{2}{c}{$q=0.75$} \\
\colhead{Source} & \colhead{A} & \colhead{C} &  \colhead{A} & \colhead{C} 
}
\startdata
$\bar{\chi}^2_{min}$ & 1.84 & 2.46 & 1.86 & 2.50 \\
$n_0$ ($\mathrm{cm^{-3}}$) & 2.70 $\pm$ 0.28 & 2.00 $\pm$ 0.14 & 2.74 $\pm$ 0.25 & 1.95 $\pm$ 0.14 \\
$T_0$ ($10^2$~K) & 1.97 $\pm$ 0.14 & 1.72 $\pm$ 0.09 & 2.05 $\pm$ 0.13 & 1.92 $\pm$ 0.08 \\
$b$ (MHz) & 1.98 $\pm$ 0.05 & 2.68 $\pm$ 0.06 & 2.00 $\pm$ 0.05 & 2.70 $\pm$ 0.06 \\
$R_{\mathrm{CH_3CN}}$ ($10^2$~AU) & 7.16 $\pm$ 0.27 & 9.01 $\pm$ 0.30 & 7.13 $\pm$ 0.18 & 8.88 $\pm$ 0.24 \\
\sidehead{Parameters set by the continuum models:}
$p$ & 1.52 & 1.52 & 1.15 & 1.14 \\
$R_{\mathrm{cav}}$ ($10^2$~AU) & 2.48 & 2.17 & 2.50 & 2.18 \\ 
\tableline\tableline
\sidehead{Parameters derived from the models:}
$b$ ($\mathrm{km \, s^{-1}}$) \tablenotemark{a} & 
    2.69 $\pm$ 0.07 & 3.64 $\pm$ 0.08 & 2.72 $\pm$ 0.07 & 3.67 $\pm$ 0.08 \\
$M_{\mathrm{vir}}$ (\Msun) & 11.7 & 26.9 & 13.1 & 29.7 \\
$L$ ($10^{4}$\Lsun) & 1.5 & 0.9 & 1.8 & 1.4 \\
$L/M_g$ ($10^3 \; \mathrm{L_{\sun}/M_{\sun}}$)  & 3.5 & 2.6 & 3.2 & 3.4 
\enddata
\tablenotetext{a}{Velocity line widths are calculated from the frequency line widths.}
\end{deluxetable}

\subsection{Thermal Coupling between the Dust and Gas Components \label{sec_coupling}}
The dust warmed by radiation from embedded sources determines the thermal 
structure of hot cores.
Since line heating by atoms and molecules is significantly weaker than the continuum 
absorption by the dust, the gas absorbs very little of the stellar radiation and is further 
shielded by the dust.
Therefore, the dust is hotter than the gas and is the dominant heating agent for the gas.
Two possible heating processes transfer energy to the gas: collisions with dust grains 
and line absorption of the IR radiation field.
Our goal here is to show whether thermal coupling between the gas and dust components 
can be achieved in hot cores rather than to make detailed calculations.

The thermal emission of the warm dust builds an intensive IR radiation field characterized 
by a radiation temperature equal to the dust temperature, $T_d$.
The molecular hydrogen has a great number of vibration-rotation transitions 
({\it e.g.} fundamental $v = 0 \rightarrow 1$ lines at 2.4~$\mu$m) and pure rotational lines 
({\it e.g.} $J = 2 \rightarrow 0$ at 28~$\mu$m) in the IR band that can respond to the 
ambient IR radiation field.
If the optical depths of these transitions are large in the hot core, the hydrogen excitation 
temperature should approach the radiation temperature $T_d$ at all time.
Then collisions between molecules will convert the internal energy to (macroscopic) 
kinetic energy.
On the other hand, if the optical depths are small, line heating will be less important 
for warming up the gas.
We have estimated the optical depths for the transitions of 
$(vJ \rightarrow v^{\prime}J^{\prime}) = (11 \rightarrow 01)$ at 2~$\mu$m and 
$v=0$ $J = 2 \rightarrow 0$ at 28~$\mu$m.
Given a minimum line width of $2.7 \; \mathrm{km \, s^{-1}}$ and a maximum column density 
of $N_{\mathrm{H_2}} \simeq 4.4 \times 10^{24} \; \mathrm{cm^{-2}}$ in our models, 
the maximum optical depth of H$_2$ transitions in the bands of 2~$\mu$m and 28~$\mu$m 
is about 1.
The optical depths of H$_2$ transitions are too small to ensure good thermal coupling 
through line heating.

Dust-grain collisions are believed to be an important heating process to explain 
the gas temperature of the molecular clouds with embedded sources 
\markcite{Goldreich74,Takahashi83}({Goldreich} \& {Kwan} 1974; {Takahashi}, {Silk}, \&  {Hollenbach} 1983).
The heating rate per H$_2$ molecule by collisions with dust grains is given 
by \markcite{Takahashi83}{Takahashi} {et~al.} (1983) as
\begin{eqnarray}
\frac{\Gamma_{\mathrm{grain}}}{n_{\mathrm{H_2}}} 
  &=& n_{\mathrm{H}} \, \Sigma_d \, \bar{\upsilon} \, \alpha \, 
           2k(T_d - T_g) \\
  & \simeq & 1.4 \times 10^{-32} \, n_{\mathrm{H}} \, (T_d - T_g) 
          \left( \frac{\alpha}{0.5} \right) \left( \frac{\bar{\upsilon}}{1\;\mathrm{km \, s^{-1}}} \right)
          \nonumber \\ 
  & &  \left( \frac{\Sigma_d}{1.0 \times 10^{-21} \; \mathrm{cm^2}} \right)
          \; \mathrm{erg \, s^{-1} \, {H_2}^{-1}}
\end{eqnarray}
where $n_{\mathrm{H}}$ is the number density of hydrogen nuclei, 
$\Sigma_d = n_d \sigma_d/n_\mathrm{H}$ is the
projected grain area per H nucleus, and $\alpha$ is the accommodation coefficient 
of the grain surface for an H$_2$ molecule \markcite{Burke83}({Burke} \& {Hollenbach} 1983).
The mean speed of H$_2$, $\bar{\upsilon}= (8 k T_g / \pi m_{\mathrm{H_2}})^{1/2}$,  
is about $1 \; \mathrm{km \, s^{-1}}$ at gas temperature $T_g \simeq 100 \; \mathrm{K}$.
Although the mean speed increases with the gas temperature, the value should be 
good for an order-of-magnitude estimate.
The gas heating rate is roughly equal to 
$\Gamma_{\mathrm{grain}}/n_{\mathrm{H_2}} = d[\case{5}{2} k (T_g - T_d)]/dt$.
Therefore, the gas-dust relaxation time, $t_{gd}$, is approximately
\begin{equation}
t_{gd} = \frac{2.5 \times 10^{16}}{n_{\mathrm{H}}} \; \mathrm{s}.
\end{equation}
Given the minimum dust density of $\rho_0 = 2.48 \times 10^{-18} \; \mathrm{g \, cm^{-3}}$ 
(see Table~\ref{pflist}) in our models, the corresponding number density of H nuclei is 
$n_{\mathrm{H}} = 1.5 \times 10^{8} \; \mathrm{cm^{-3}}$ if a gas-to-dust ratio of 100 is assumed.
Therefore, the gas-dust relaxation time is about
$t_{gd} \lesssim 1.7 \times 10^{8} \; \mathrm{s} = 5.4 \; \mathrm{yr}$, which is significantly 
shorter than the lifetime of hot cores.
As long as the stellar radiation does not vary greatly on the time scale of the gas-dust 
relaxation time, the gas and dust components are well coupled thermally by collisions.

\subsection{The Nature of the Velocity Gradient}
\begin{figure}
\plotone{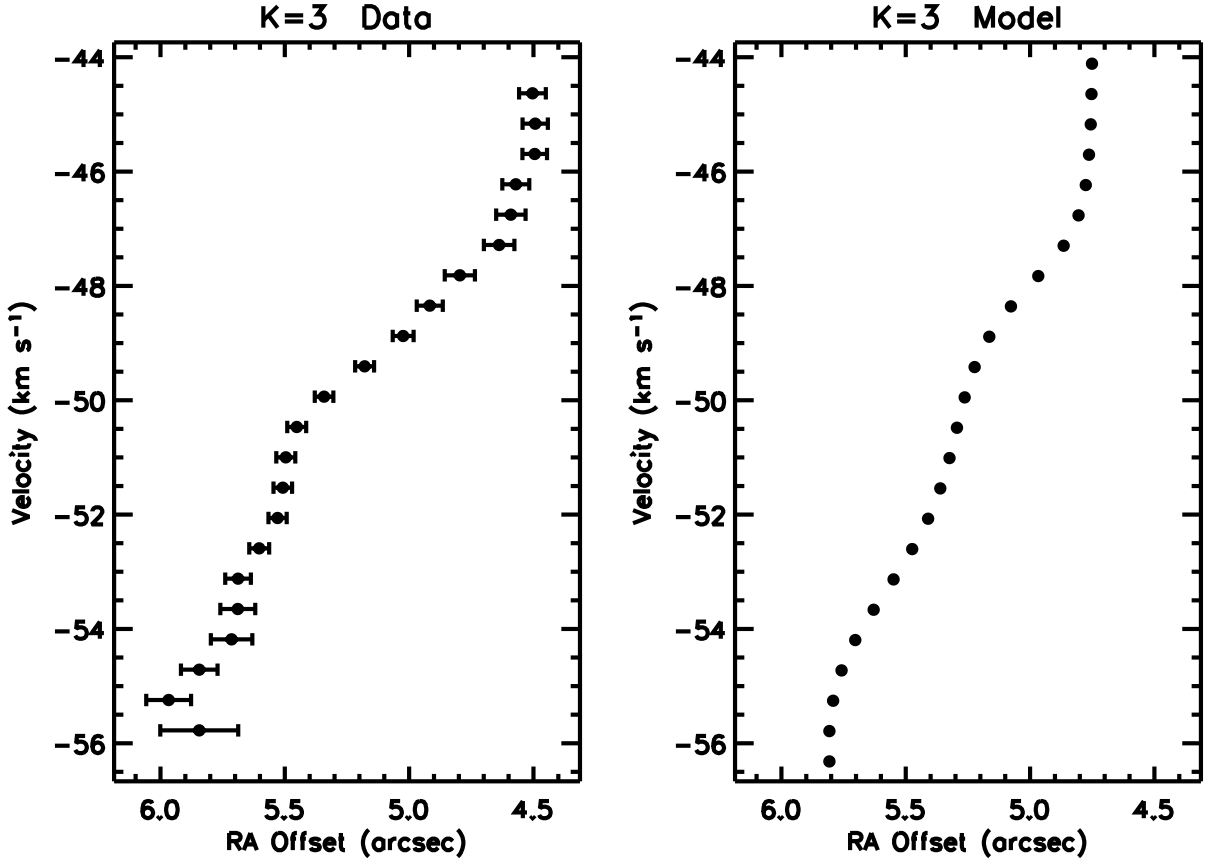}
\caption{Right ascension offset of the Gaussian peak position versus 
the radial velocity in W3(H$_2$O).
A velocity gradient is clearly seen in the observed $K=3$ component (left) 
and the model images (right).
The velocity gradient generally agrees with the result in \markcite{Wyrowski97}{Wyrowski} {et~al.} (1997).  
The discontinuity of the position drift on the edges and the twitch at offset 5\farcs4 suggest 
the presence of multiple clumps. \label{K3vpfit}}
\end{figure}

The linear velocity gradient suggested by \markcite{Wyrowski97}{Wyrowski} {et~al.} (1997) is also detected in 
our methyl cyanide lines.
For the purpose of easy comparison, we only perform the Gaussian fits on 
the $K=3$ component and the result is shown in the left panel of Fig.~\ref{K3vpfit}.
The position uncertainty, $\sigma_{\theta}$, of a Gaussian emission 
feature due to thermal noise is given by \markcite{Reid88}{Reid} {et~al.} (1988) as
\begin{equation}
  \sigma_{\theta} = \left( \frac{4}{\pi} \right)^{1/4} 
  \frac{\theta_{\,\mathrm{FWHM}}}{\sqrt{8 \ln2}} \;
  \frac{1}{\mathrm{SNR}} = 0.45 \; \frac{\theta_{\,\mathrm{FWHM}}}{\mathrm{SNR}},
\end{equation}
where $\theta_{\,\mathrm{FWHM}}$ is the full-width at half-maximum of the emission 
feature, and SNR is the signal-to-noise ratio, which is set equal to the peak intensity 
divided by the rms noise well away from the emission.

\begin{figure}
\plottwo{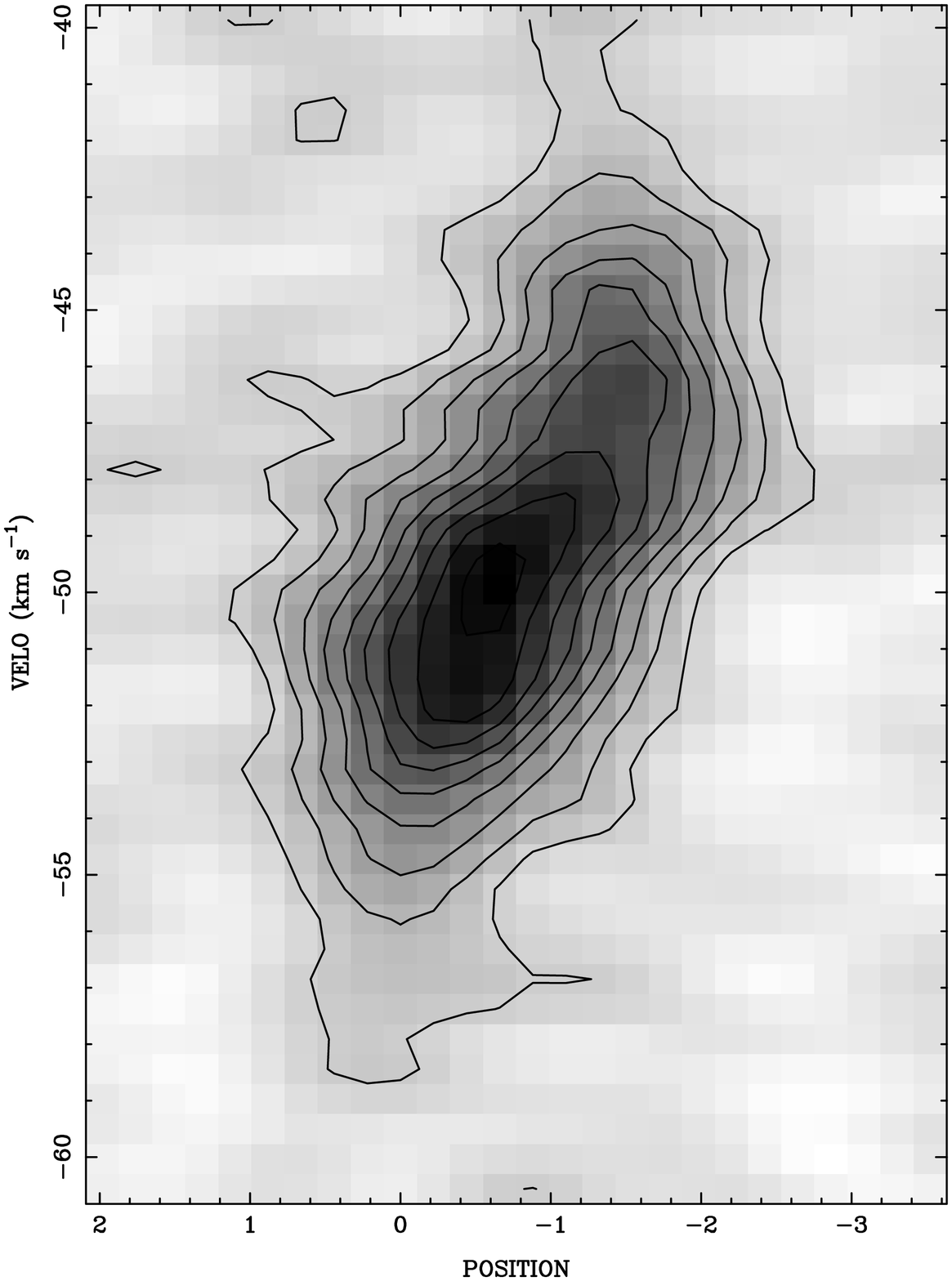}{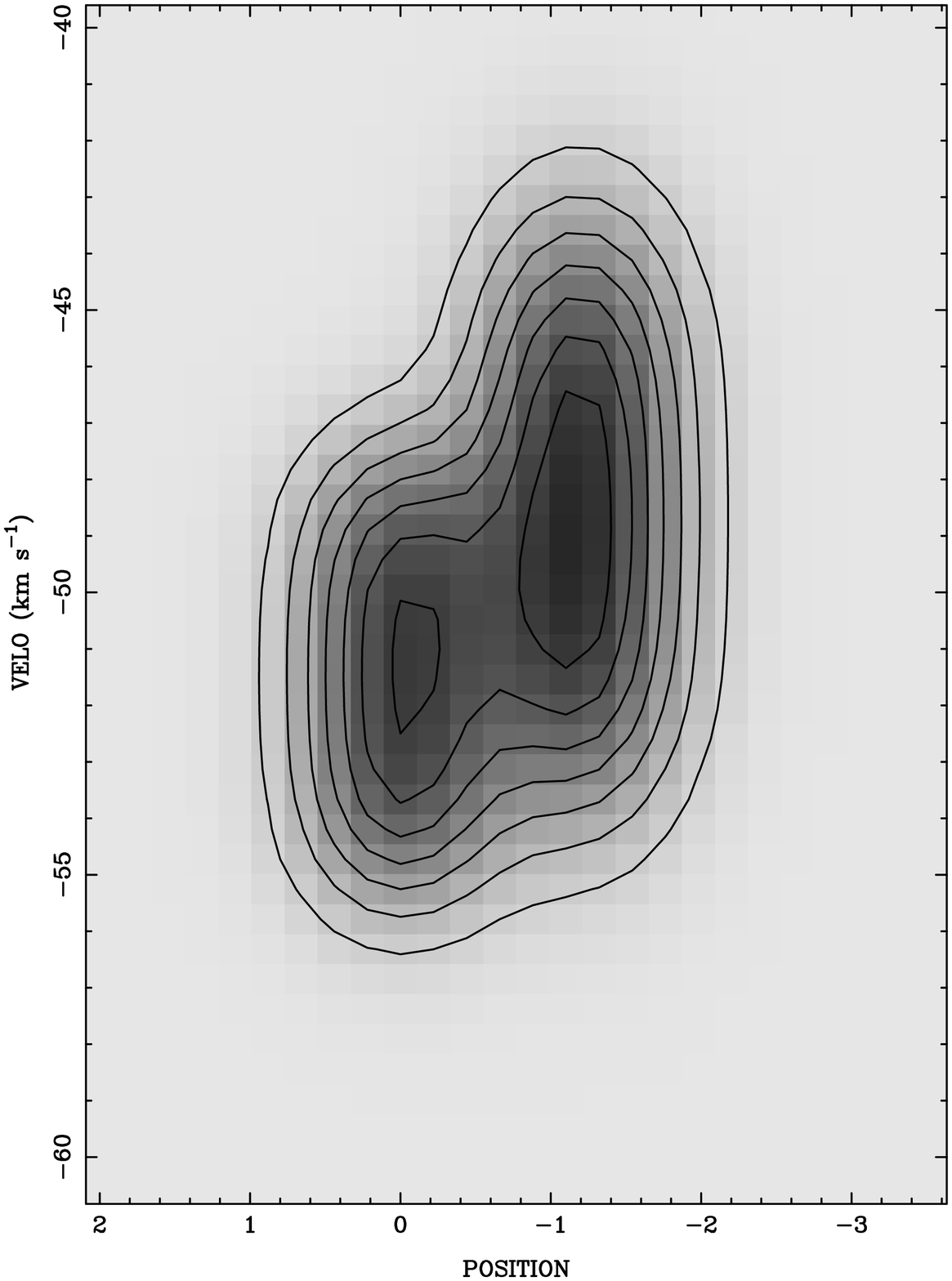}
\caption{Velocity-position plot of the $K = 3$ component of 
the $\mathrm{CH_3CN} \;\; J = 12-11$ transition (left) and the corresponding plot 
for the model data cube (right).  Contour levels correspond to 5.7 to 51.3 by 5.7 
($2 \sigma$)~K. \label{K3vp}}
\end{figure}

A similar velocity gradient along the East-West direction across \wtw\ is clearly seen. 
With improved angular resolution, our data are able to trace over a broader velocity 
range with better sampling and provide a few important details.  
Our position drift shows turnovers, where the peak positions stay 
unchanged at the two ends, and a twitch, where the velocity has a larger increment 
(from $-52 \; \mathrm{km\,s^{-1}}$ to $-50.5 \; \mathrm{km\,s^{-1}}$)
at 5\farcs5 east to the phase center.
These irregular behaviors in the velocity gradient can result from two unresolved 
clumps with a small radial velocity difference.  
In order to inspect this hypothesis, we have generated a binary data cube by 
adding linearly the methyl cyanide models of the two sources and performed 
the same Gaussian fitting to the model images.
The model images show similar behaviors (right panel of Fig.~\ref{K3vpfit}) and 
suggest that the binary rotation can explain the irregularities in the velocity gradient.
To have a better understanding, we have taken a velocity-position slice that passes
through the two continuum peaks (Fig.~\ref{K3vp}).
At first sight, a velocity gradient seems obvious; however, the sharp edges suggest that 
the velocity gradient is a combination of two unresolved clumps.
The same position-velocity slice with the model data cube shows that two connected ellipses 
are expected from a partially resolved binary motion.

Our angular resolution of 1\arcsec\ has improved by nearly a factor of 2 compared with 
the previous study \markcite{Wyrowski97}({Wyrowski} {et~al.} 1997).
This better angular resolution has provided important clues to reveal the rotation of 
an unresolved binary.
When the emission is only resolved in one direction, a Gaussian function can be a misleading approximation for the emission morphology.
We conclude that the linear velocity gradient is actually a result of unresolved binary rotation 
rather than the Keplerian rotation of a disk or the expansion of a bipolar outflow.

\section{SUMMARY}
\begin{enumerate}
\item 
The \wtw\ hot core is the best case to date that shows a massive protobinary
system at such an early evolutionary stage that it is still associated with nebular emission 
from the surrounding dust cocoon.
This hot core is also a good example showing fragmentation as part of the star 
formation process for massive stars.
\item
We have obtained the best angular resolution that has ever been observed 
for \wtw\ at millimeter wavelengths: the synthesized beam size reaches 
0\farcs26 at 1.4~mm and 0\farcs4 at 2.8~mm.
Our continuum maps show a double-peaked morphology, suggesting an embedded 
protobinary system with a projected separation of $2.43 \times 10^3 \; \mathrm{AU}$.
\item
The kinematics gives a radial velocity difference of $2.8 \; \mathrm{km \, s^{-1}}$, 
corresponding to a minimum binary mass of 22\Msun.   
The continuum models provide a nebular mass of 4 to 6\Msun\ for 
source~A and 3 to 4\Msun\ for source~C.
\item
The two binary members have a dust opacity law of $\kappa_{\nu} \propto \nu$,
suggesting that grain growth has begun.
\item 
If the temperature profile follows $T \propto r^{-0.4}$, the density distribution 
for each source has a power-law index of $-1.52$, which is close to the 
power-law index of $-1.5$ for free-fall collapse.
\item
The luminosity of source~A is in the range of $(1.5 - 1.8) \times 10^4$\Lsun\ and 
that of source~C is in the range of $(0.9 - 1.4) \times 10^4$\Lsun.
The luminosities of the two binary members correspond to ZAMS 
spectral type B0.5 - B0. 
Since a star of spectral type earlier than B3 will produce an observable \HII\ region, 
we expect each member of the binary to develop its own \HII\ region in the future.
\item
Based on the brightness temperature argument of the five $K$ components, 
the methyl cyanide emission should emerge from regions significantly smaller 
than the extent of the warm dust.
The methyl cyanide line model provides a temperature estimate of $200$~K 
for source~A and $182$~K for source~C at the reference radius of 500 AU.
\item
Excluding any mass contained in central protostars, the derived luminosity to 
nebular mass ratio, $L/M_g$, is larger than $700 \; \mathrm{L_{\sun}/M_{\sun}}$, 
but smaller than the Eddington limit by one order of magnitude.
\item
The virial masses derived from the line widths set upper limits for the masses at 
12\Msun\ for source~A and 28\Msun\ for source~C.
\end{enumerate}
 
We wish to thank Mel Wright for help with the BIMA data analysis and plotting 
programs, MIRIAD and WIP.  Thanks to Dick Plambeck for help with 
many questions that arose during this research.
Thanks also due to Al Glassgold for the discussion on the thermal
coupling between the dust and the gas.
This work was supported in part by NSF grant AST93-14847.

\appendix
\section{PROPERTIES OF METHYL CYANIDE MOLECULES \label{sec_pmc}}
\begin{figure}
\plotone{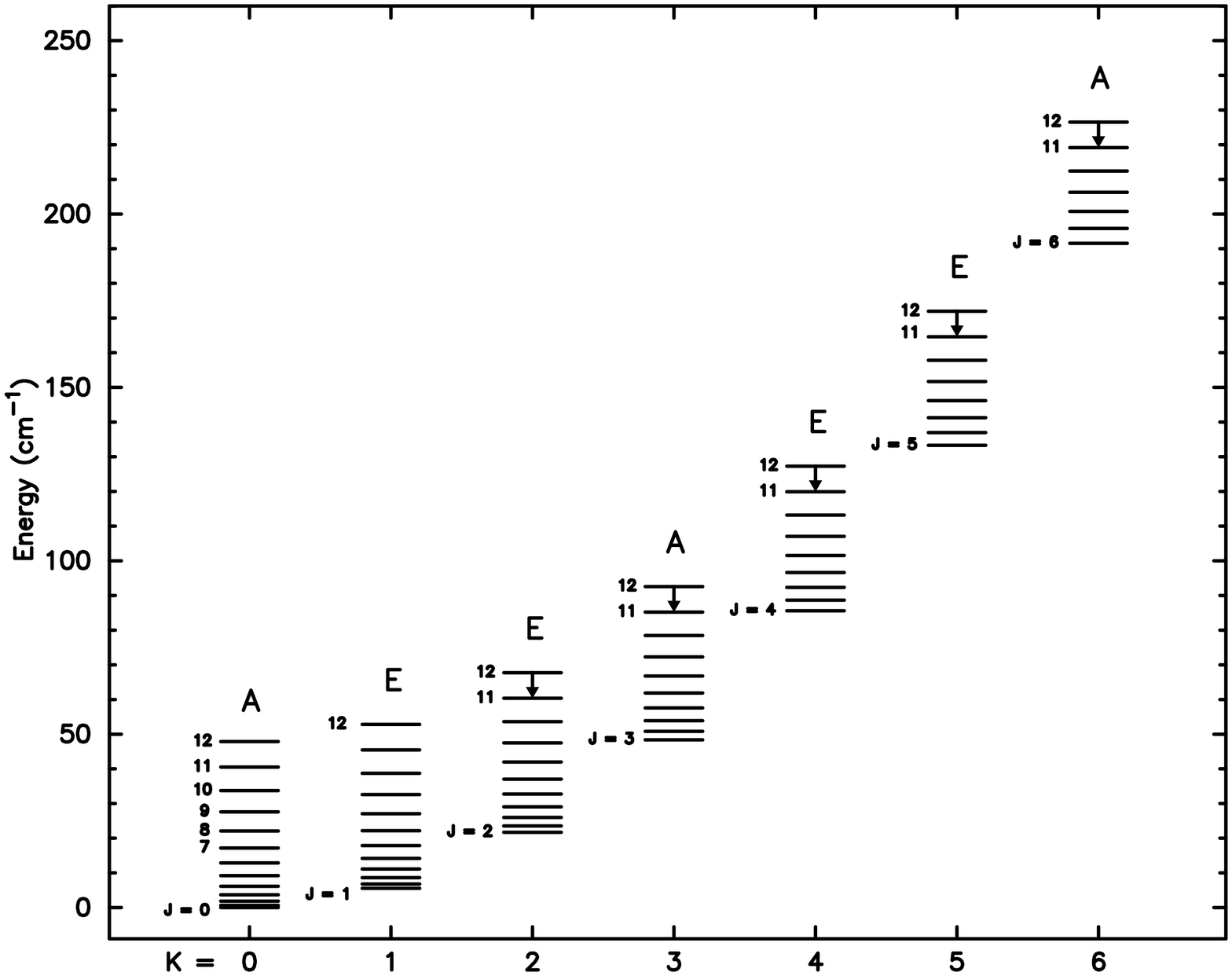}
\caption{Energy level diagram of $\mathrm{CH_3CN}$ for $J$ levels up to
$J=12$ and $K$ ladders from $K=0$ through $K=6$.
The observed transitions are indicated by arrows.
The A and E letters label the two independent species of $K$ ladders 
distinguished by symmetry.  \label{level}}
\end{figure}

The energy levels of a symmetric-top molecule can be described by two quantum 
numbers, $J$ and $K$, where $J$ is the total angular momentum and $K$ is
its projection on the symmetry axis of the molecule.
The rotational energy $E_{JK}$ of a symmetric-top rotor including centrifugal
stretching can be written as \markcite{Sutton86}({Sutton} {et~al.} 1986; \markcite{Anttila93}{Anttila} {et~al.} 1993)
\begin{eqnarray}
\frac{E_{JK}}{h} &=& B \, J (J+1) + (A-B) \, K^2  \nonumber \\ 
                          & & - D_J \, J^2 (J+1)^2 - D_{JK} \, J(J+1) K^2 - D_K \, K^4  
\end{eqnarray}
where $A$, $B$, $D_J$, $D_{JK}$, and $D_K$
are molecular constants and we have adopted the values given by 
\markcite{Pearson96}{Pearson} \& {Mueller} (1996).
The energy level diagram of methyl cyanide is showed in Fig.~\ref{level}.
Since a symmetric-top molecule has no dipole moment perpendicular to the
symmetry axis, a radiative transition can not change the angular momentum 
along this axis.
Therefore, radiative transitions are only allowed between adjacent $J$ levels
within a single $K$ ladder.
The spacing between adjacent $J$ levels only decreases slightly in different
$K$ ladders.
As a result, the transitions of adjacent $J$ levels, {\it e.g.} $J \rightarrow J-1$, 
in various $K$ ladders happen at nearly the same frequency and can
be observed simultaneously.
Each successive $K$ component is shifted to a slightly lower frequency
due to the centrifugal distortion.

The three identical hydrogen nuclei in methyl cyanide create threefold symmetry.
When a pair of hydrogen nuclei are exchanged, the combined effect of 120$^\circ$~rotation 
and nuclear spins brings additional symmetry requirements that further divide the molecules 
into two distinct species: ortho- and para-$\mathrm{CH_3CN}$. 
Rotational levels with $K = 3n$ ($n$ is an integer), belong to the ortho-species, while 
levels with $K = 3n \pm 1$ belong to the para-species.
The ortho- and para-species are sometimes referred to as the A and E species, respectively.
The A and E species are essentially independent since the probability of changing 
nuclear spin state in a collision is very small and radiative transition between 
species is forbidden. 
Because of the symmetry requirement, the statistical weight, $g_{IK}$, of the A species 
is twice larger than that of the E species with the exception of the $K = 0$ ladder, 
where half of the levels are forbidden due to the total cancellation of the wave function.

The collision rate coefficients for downward transitions, $K_{ul}$, of methyl cyanide were first calculated by \markcite{Cummins83}{Cummins} {et~al.} (1983) and expanded to higher energy levels and higher temperature 
up to $140$~K by \markcite{Green86}{Green} (1986) to better match the conditions in star-forming regions.  
The basic idea is to express collision rate coefficients in terms of a few fundamental rates,
$Q(L, M)$, that are simply related to excitation out of the lowest $J=K=0$ level.
The collision rate coefficient for the transition $(J,K) \rightarrow (J',K')$ of 
a symmetric-top molecular can be expressed as
\begin{eqnarray}
    K_{ul} &=& (2J'+1) \sum_{L=|J-J'|}^{J+J'} 
               \left( \begin{array}{ccc} 
                       J' & L & J \\ 
                       K' & K - K' & -K  
               \end{array} \right)^2 Q(L, |K - K'|) \quad \mathrm{cm^3 \, s^{-1}}, 
               \quad K'=0;  \label{Kul_0} \\
K_{ul} &=& (2J'+1) \sum_{L=|J-J'|}^{J+J'} \left[
                    \left( \begin{array}{ccc}  J' & L & J \\ K' & K - K' & -K \end{array} \right)^2 
                    Q(L, |K - K'|) \right. \nonumber \\
          & & \hspace{2.5cm} + \left. 
             \left( \begin{array}{ccc}  J' & L & J \\ -K' & K + K' & -K \end{array} \right)^2 Q(L, K + K') 
             \right] \quad \mathrm{cm^3 \, s^{-1}}, \quad K' \neq 0, \label{Kul_n0} 
\end{eqnarray}
where the large bracket is a 3-$j$ symbol.
The second term in the large square bracket on the right-hand side of Eq.~(\ref{Kul_n0}) is 
much smaller than the first term and was neglected in our calculations in Sect.~\ref{sec_tlp}.
The fundamental rates, $Q(L, M)$, that we used in this study are as given by \markcite{Green86}{Green} (1986).

\section{VIRIAL MASS FOR A POWER-LAW DENSITY DISTRIBUTION \label{sec_Mvir}}
Given a power-law density distribution of $\rho = \rho_0 (r/r_0)^{-p}$,
the enclosed mass within a radius of $r$ is $m(r) = 4 \pi \, \rho_0 \, r_0^p \, r^{3-p} /(3-p)$.
The gravitational potential energy, $U$, for a clump of radius $R$ can be calculated by
\begin{eqnarray}
U &=& -4\pi \, G \, \int_0^R m(r) \, \rho(r) \, r dr  \nonumber \\
    &=& -\frac{16\pi^2 \, G \, \rho_0^2 \, r_0^{2p}}{(3-p)(5-2p)} \, R^{5-2p} \nonumber \\\
    &=& \frac{3-p}{5-2p} \frac{G M^2}{R}, \\
\end{eqnarray}
where $M = m(R) = 4\pi \, \rho_0 \, r_0^p \, R^{3-p}/(3-p)$ is the total mass of the clump.
The kinetic energy can be written as $K = 3NkT/2 = 3 M b^2/4$, where $N$ is the total
number of particles, and $b$ is the gaussian linewidth in Eq.~(\ref{eq_phi}).
Assuming the gas is in quasistatic equilibrium, the gravitational potential energy of 
the clump can be related to its kinetic energy by the virial theorem, $U=-2K$.
Therefore, the virial mass can be written as
\begin{equation}
M = \frac{3}{2} \, \left(\frac{5-2p}{3-p}\right) \left(\frac{b^2 \, R}{G}\right).
\end{equation}


\end{document}